\documentclass[journal]{IEEEtran}
\usepackage{cite}
\usepackage{amsmath,amssymb,amsfonts}
\usepackage{algorithmic}
\usepackage{graphicx}
\usepackage{textcomp}
\usepackage{enumitem}
\usepackage[caption=false]{subfig}
\usepackage{color,soul}

\begin{document}

\title{Reconfigurable Intelligent Surfaces for 6G Mobile Networks: An Industry R\&D Perspective}

\author{
    \IEEEauthorblockN{
        Maik Sode\IEEEauthorrefmark{1}, Michael Ponschab\IEEEauthorrefmark{1}, Lucas N. Ribeiro\IEEEauthorrefmark{1}, Sven Haesloop\IEEEauthorrefmark{2}, Ehsan Tohidi\IEEEauthorrefmark{2}\IEEEauthorrefmark{3}, Michael Peter\IEEEauthorrefmark{2}, S\l{}awomir~Sta\'nczak\IEEEauthorrefmark{2}\IEEEauthorrefmark{3}, Bilal H. Mohamed\IEEEauthorrefmark{4}, Wilhelm Keusgen\IEEEauthorrefmark{4}, Heinz Mellein\IEEEauthorrefmark{5}, Eslam Yassin\IEEEauthorrefmark{6}, Bernd Schroeder\IEEEauthorrefmark{6}
    }
    
    \IEEEauthorblockA{
      \vspace{0.4cm}
        \IEEEauthorrefmark{1}Ericsson Antenna Technology Germany GmbH, 83026 Rosenheim, Germany\\
        \IEEEauthorrefmark{2}Fraunhofer Heinrich-Hertz-Institut, 10587 Berlin, Germany\\
        \IEEEauthorrefmark{3}Network Information Theory, Technische Universit\"at Berlin, 10623 Berlin, Germany \\        
        \IEEEauthorrefmark{4}Department of High Frequency Systems, Technische Universit\"at Berlin, 10623 Berlin, Germany \\
        \IEEEauthorrefmark{5}Rohde \& Schwarz GmbH \& Company KG, 81671 Munich, Germany\\
        \IEEEauthorrefmark{6}brown-iposs GmbH, 53229 Bonn, Germany\\
        \thanks{This work was supported in part by the Federal Ministry of Education and Research (BMBF) of the German Government under the ``6G LICRIS'' project.}
        \thanks{Corresponding author: Michael Ponschab (e-mail: michael.ponschab@ericsson.com).}      
        \thanks{ } 
    }
}

\maketitle

\IEEEpubid{\begin{minipage}{\textwidth}\ \\[14pt] \centering
  DOI: 10.1109/ACCESS.2024.3485227 \copyright 2024 IEEE. Personal use is permitted, but republication/redistribution requires IEEE permission.\\
  See http://www.ieee.org/publications standards/publications/rights/index.html for more information.
\end{minipage}}

\begin{abstract}
    The reconfigurable intelligent surface (RIS) technology is a potential solution to enhance network capacity and coverage without significant investment in additional infrastructure in 6G networks. This work highlights the interest of the mobile communication industry in RIS, and discusses the development of liquid crystal-based RIS for improved energy efficiency and coverage in the millimeter-wave band. Furthermore, the paper discusses perspectives and insights from an industry R\&D point of view, addressing relevant use cases, technical requirements, implementation challenges, and practical considerations for RIS deployment optimization in the context of 6G networks. A hardware design of an RIS with liquid crystal at 28~GHz is presented. A propagation model for RIS as a new part of the system architecture is discussed, with approaches of semi-empirical models, geometric models, and their combination through the application of artificial intelligence/machine learning. Finally, a channel model for deployment optimization and dimensioning is presented, with the findings that a rather large RIS is favorable for coverage improvement, as well as greater attenuation at higher frequencies combined with a smaller RIS size. 
\end{abstract}

\begin{IEEEkeywords}
    6G, reconfigurable intelligent surfaces, network, liquid crystal, unit cell, network densification, coverage issues, channel models, propagation modeling, over-the-air (OTA) -testing.
\end{IEEEkeywords}



\section{Introduction}
\label{sec:introduction}

\IEEEPARstart{T}{he} fifth-generation (5G) mobile networks are currently being rolled out worldwide. Theoretically, 5G should provide data rates of up to 10 Gbps on average and latencies below 10 ms. However, it is clear that with future applications, for example in connection with telepresence, mixed realities, or for highly specialized use in an industrial context, the requirements for data rate, reliability, connection density, and latency will continue to increase. Among other targets, the sixth-generation (6G) aims at larger transmission rates, lower latencies, and more energy efficiency. It provides connectivity even under unfavorable conditions to enable safety-critical applications such as the connection for unmanned autonomous vehicles. These targets cannot be achieved by simply deploying additional base stations, but require alternative, innovative solutions. With regard to the long-term national and international climate goals, as well as the social acceptance of 6G, it is crucial to focus on energy efficiency in all necessary developments and network expansions. Sustainability is highlighted as an essential criterion on which 6G should build upon in comparison to~5G.

Potential solutions to achieve 6G's targets include utilizing new spectrum and implementing further network densification with distributed Multiple-Input Multiple-Output (MIMO) systems concepts~\cite{jiang2021road}. However, these measures result in an increase in energy requirements, which can only be mitigated by developing more energy-efficient components. Transitioning to higher frequencies also leads to increased propagation losses, requiring extremely dense deployments with direct line of sight (LOS). Additionally, existing networks may not easily support future applications with different radio coverage requirements, such as aerial connectivity, necessitating network coverage extension to the lower airspace. Cooperative relaying systems~\cite{hammerstrom2007power} have also been proposed as a potential solution to improve coverage and have even been incorporated in Long-term Evolution (LTE) Advanced~\cite{tran2012overview}. However, they require reliable backhaul connection, which increases its implementation complexity and cost.

A novel approach to enhance network capacity, coverage, and range without additional base stations or repeaters is through reconfigurable intelligent surfaces (RIS)~\cite{di2020smart,tang2021}. RIS involves using passive controllable surfaces to shape the radio environment, influencing wave propagation for various purposes such as improving cell coverage and controlling interference. RIS can also enhance spatial diversity and channel capacity in MIMO systems and find potential applications in emerging technologies such as the Internet of Things (IoT), Industry~4.0, ground-to-air communications, localization, sensing, and physical layer security\cite{di2020smart}. Due to its potential to significantly improve the performance of radio access networks, RISs have caught the attention of academia, industry, funding bodies, and regulators. Consequently, significant investments have been directed towards the development and evaluation of this technology for 6G. Industry players such as ZTE~\cite{zte24,zhao2023zte,liu22zte}, AGC~\cite{agc21}, SK Telecom~\cite{sk23}, and NTT~\cite{ntt20} have started to invest in the research and development of RIS to expand their product portfolios.

For example, the 6G-LICRIS (Liquid Crystal Reconfigurable Intelligent Surfaces) project aims to develop RIS based on liquid crystal (LC) for surface reconfigurability from a holistic network perspective. Compared to semiconductor-based RIS, LC is more energy efficient in higher frequencies, scalable, cost-effective, and a promising technology candidate for RIS hardware implementation. The aim of the 6G-LICRIS project is to develop reconfigurable intelligent surfaces and integrate them into the mobile network. To achieve this, new materials based on liquid crystals are being developed. The newly developed technology will be integrated into a test mobile network to study radio wave propagation using the RIS. Finally, the integration of RIS into a real radio environment to demonstrate end-to-end data transmission in the actual mobile network is considered~\cite{vdivde}.

The project has identified several key steps to improve its features and categorizations. It will follow the European Telecommunications Standards Institute (ETSI) reports and use metamaterial and reflectarray structures. The RIS will have a passive design, reflecting the outgoing RF wave and saving power compared to base station antennas. It will operate in the mmWave frequency range with time division duplex (TDD) duplex mode. The RIS's goal is to enhance coverage, spectral efficiency, and beam management while improving energy efficiency. The RIS's deployment scenario can be static in both indoor and outdoor environments. It can have a centralized or distributed management system with a network-controlled mode of operation. The communication and networking topology can be both indoor and outdoor, and the RIS should focus on improving indoor signal strength from an outdoor base station. While the RIS will not have sensing or localization functionalities, it should make use of channel estimation and precoding offboard. Key performance indicators should include rate or throughput, coverage, and energy efficiency, which depend on design parameters like size, carrier frequency, and bandwidth. The connectivity and reliability through deployment scenarios should be improved to utilize single or multiple RISs controlled individually to maximize coverage.

The objective of this article is to share the authors' perspective and learnings on RIS coming from an industry research and development (R\&D) point of view at the beginning of the 6G-LICRIS project. More specifically, relevant use cases for future investigation, their technical requirements, and existing challenges are discussed in Section~\ref{sec:usecases}. Similar considerations are presented in~\cite{li2023considerations, aastrom2024ris}, but the present work differentiates itself from the existing literature by diving deeper into implementation and deployment aspects. In particular, the hardware implementation of novel LC-based concepts are presented in Section~\ref{sec:hw}. The practical problem of channel modeling, RIS deployment optimization and dimensioning are discussed in Section~\ref{sec:depl}. These problems are discussed with novel insights into the line-of-sight coverage improvement of RIS in urban deployment and the corresponding RIS size optimization. Finally, the paper is concluded with potential topics for further investigation that shall support the incorporation of RIS in mobile communication standards.

\section{Relevant Use Cases, Technical Requirements and Challenges}\label{sec:usecases}

This section presents potential use cases for the RIS technologies and discusses fundamental technical requirements. This discussion includes regulation aspects, conformance testing points, and outlines operational requirements gaps that are not covered by existing regulation.

\subsection{Relevant Use Cases}\label{sec:usecase}

Use cases for RIS are currently being discussed in the radio communication industry, starting from coverage extension towards smart antenna applications. However, standardisation bodies, such as 3rd Generation Partnership Project (3GPP) or ETSI have just started to analyze possible opportunities. One of the first standard developing organizations (SDO) working group to look into RIS specifications and applications has been established in ETSI early in 2021, namely the ETSI ISG RIS (ETSI Industry Specification Group for RIS). Still, the technical group reports are not published and in a drafting stage. So far, they have identified the following potential use cases for RIS deployments:

\begin{enumerate}[label=(\alph*)]
    \item Signal enhancement for hidden users: The direct LOS is obstructed by a building. This reduces the signal quality on receiver side massively. By using the nearby installed RIS, a ``quasi-LOS'' connection can be established, and the signal quality is significantly improved. This application is particularly important in high-frequency scenarios (mmWave and THz communication), as these kind of signals are highly susceptible to blockage and pathloss~\cite{jiang20226g}.
    \item Interference suppression: By deploying RISs on cell edges, the desired signal strengths can be improved compared to the inter-cell interference (interference by other neighboring base stations/cells) signal strengths. This is done by passive beamforming, as the signals are fed to the target direction only. Following this the spread of signals into areas that should not actually be covered is reduced~\cite{meng2022}.
    \item Increase network capacity: The quality of signal in networks with time-varying channels can be improved~\cite{xu23} as well as in wide-band applications~\cite{an24}. For improving the channel rank conditions, RISs allow manipulation of the wireless channel by adding additional propagation paths into the desired direction. By this the channel rank can be improved~\cite{meng2022}. %
    \item Physical layer security: In addition to the above mentioned use cases, RISs can be used to improve the secrecy capacity of communication networks~\cite{makarfi2020}. The secrecy capacity is a common metric in physical layer security. It is defined as the difference between the link capacity of the legitimate transmitter/receiver and the link capacity of the eavesdropper transmitter/receiver. These link capacities are calculated based on the corresponding signal-to-noise ratios (SNR). Due to the beam shaping capability of an RIS, the SNR for the legitimate link can be improved and the SNR for eavesdropper is reduced~\cite{li2020}.
    \item Mounting of RIS on unmanned aerial vehicle (UAVs): To assist and enhance terrestrial communication, RISs can be attached to UAVs~\cite{liu2022}. This idea is similar to the mounting of RIS modules to buildings. However, attaching them to UAVs provides even more degrees of freedom with regard to positioning. It extends the coverage of the wireless signals by utilizing beamforming with UAV trajectory optimization~\cite{meng2022}.
    \item RIS to support simultaneous wireless information and power transfer (SWIPT): Internet of Things (IoT) devices like small sensors that support our daily life, are increasingly deployed worldwide. Traditionally, these devices are equipped with small batteries. One emerging approach studies the power supply of these devices by using the radiated energy for the wireless information transfer (SWIPT). By introducing RISs in such scenarios, the energy harvesting and data transmission capabilities can be boosted~\cite{meng2022,lan24,roshdy23}.
    \item Enhancement of localization and sensing capabilities using RIS: The introduction of RIS into localization and sensing applications generates new synchronization location references and configurations. Additionally, it enables new geometric measurements and improves localization accuracy~\cite{bjornson2022}.
    \item RIS-assisted Non-Orthogonal Multiple Access (NOMA): Compared to conventional Multiple Input Multiple Output (MIMO)-NOMA, RIS-assisted NOMA can improve throughput, spectral efficiency, and user connectivity by overcoming challenges such as random channel fluctuations, blockages, and user mobility~\cite{huang2022}.
    \item Enhancing indoor signal strength: Due to the intrinsic impedance difference between free-space and glass, the signal magnitude through windows is drastically reduced. To overcome this challenge, transparent reflecting intelligent surfaces, also known as transparent intelligent surfaces, provide a suitable solution for improving signal quality without greatly impairing the actual function of the glass, achieving an optical transparency of 80\%~\cite{youn2022}.
\end{enumerate}
Other applications are expected to follow, with developments in metamaterials enabling RIS design and the integration of RIS products into the ecosystem of radio communication systems by standardization bodies.

\subsection{Technical Requirements}\label{sec:requirement}

\subsubsection{Regulation}

The baseline for regulating radio equipment sold and deployed in the European marketplace, including Germany, is the European Radio Equipment Directive 2014/53/EU, commonly known as ``RED''~\cite{directive2014}.

\begin{figure}
    \centering
    \includegraphics[width=\columnwidth]{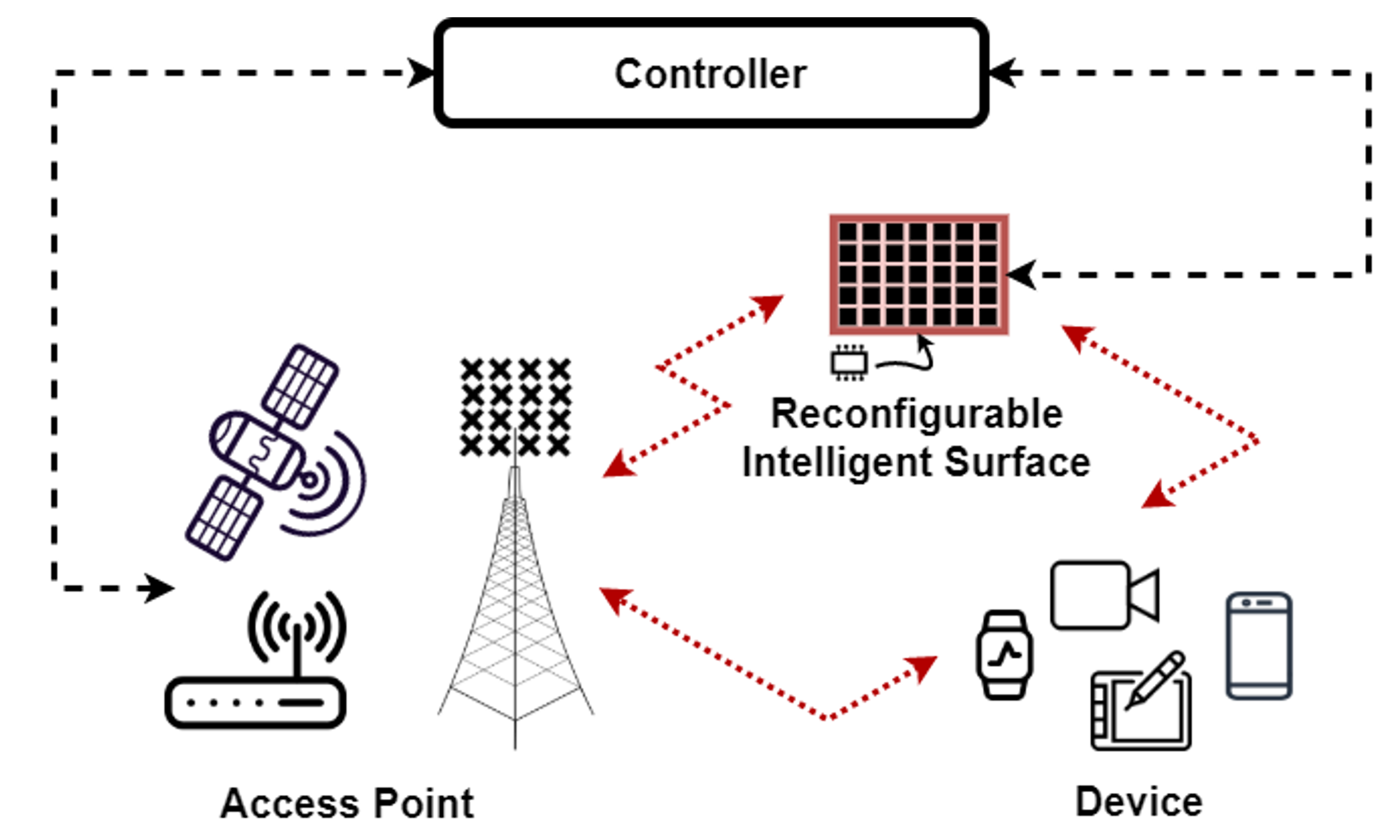}
    \caption{RIS deployment. Figure adapted from~\cite{ETSI-RIS-GR002}.}\label{fig:reg:def}
\end{figure}

As RIS modules and RIS-based products will primarily be deployed as radio network components to extend coverage, as depicted in Figure~\ref{fig:reg:def}, or offer radio beam shaping capabilities to transmitters, they must meet the regulatory requirements according to RED. Therefore, it is expected that there will be a need for RIS-specific harmonized requirements standardization in the European Union. Although there is no request from the European Commission yet, standardization bodies like ETSI have already started this process, namely the ETSI ISG RIS.

\begin{figure}
    \centering
    \includegraphics[width=\columnwidth]{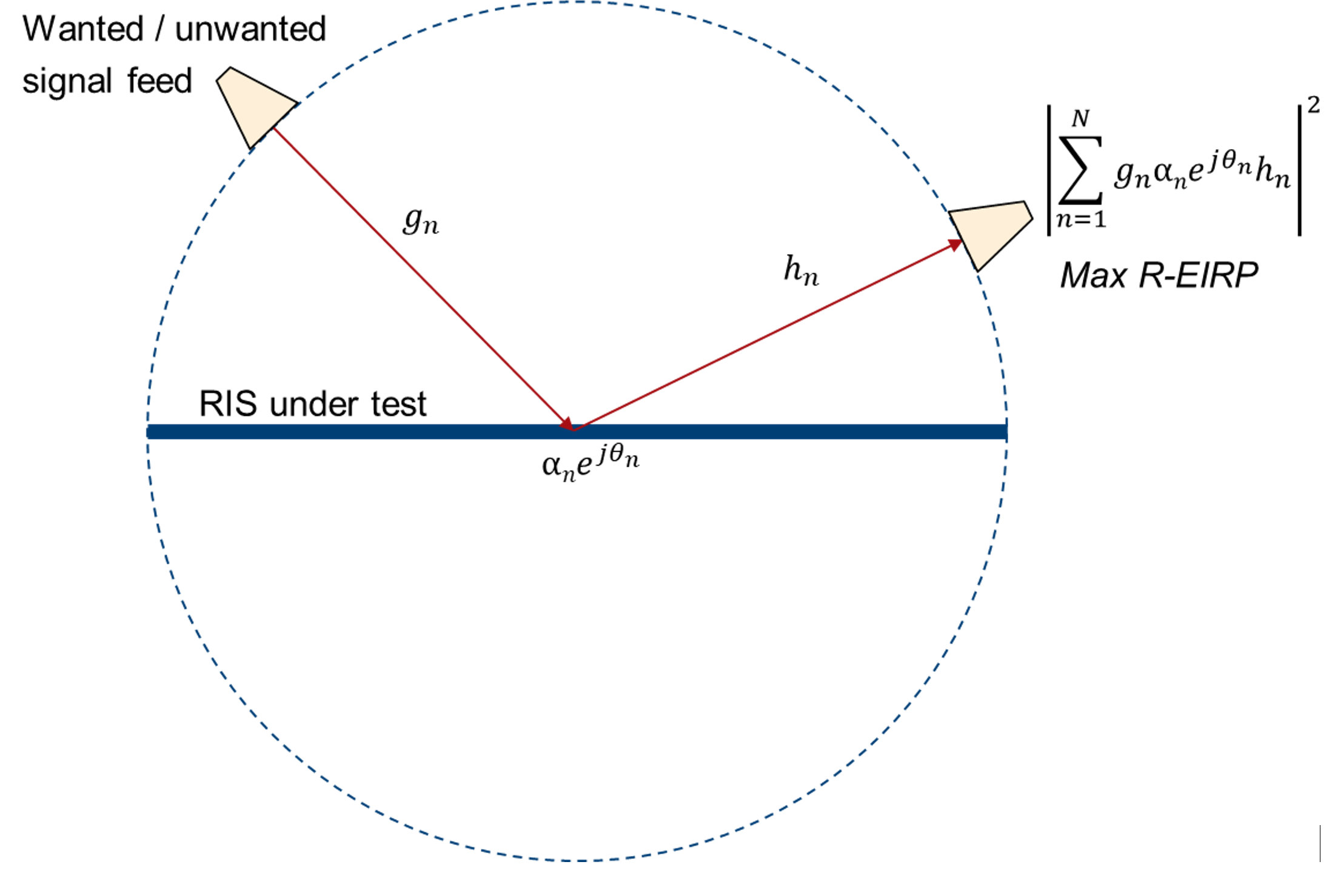}
    \caption{RIS over-the-air setup. Figure adapted from~\cite{ETSI-RIS-GR002}.}\label{fig:reg:ota}
\end{figure}

Among other stakeholders, Rohde \& Schwarz is a contributing member of the ETSI ISG RIS group, aiming to identify potential regulatory testing needs at the earliest possible stage. The common knowledge about regulatory testing will be contributed to the 6G-LICRIS project to specify potential testing requirements and adjust all RIS testing investigations accordingly. A very early outcome of the discussion on RIS testing needs in the industry has been the common conclusion toward primarily over-the-air testing in anechoic environments, similar to those known from legacy antenna test systems. Compared to antennas, where the radiated signal is typically fed via antenna connectors, the ``feed signal'' for an RIS module is the impinging radio signal. Consequently, the entire measurement setup is radiated (feed and probe signal) or ``over-the-air''. Therefore, legacy antenna measurement systems need to be extended by feed signal antennas in addition to the probe antennas, as depicted in Figure~\ref{fig:reg:ota}.

Regulatory requirements primarily focus on the protection of other radio services and the most efficient usage of the given spectrum as an important asset. Thus, it is expected that regulatory requirements will include addressing unwanted and spurious emissions caused by RIS deployment, as well as blocking and intermodulation rejection requirements.

\subsubsection{Conformance}

Conformance testing is regarded as one of the key success factors in mobile communication over the last three decades. The idea is a common industry-driven regime to test and certify mobile devices and network components, for instance, according to the common standard provided by 3GPP. Conformance testing goes far beyond regulatory testing, as its focus is on ensuring the seamless operation of devices worldwide, since radio regulations may differ across different geographical regions. As RIS might become an important component of future radio communication network architectures, RIS products will need to adhere to that conformance testing regime. Therefore, in addition to regulatory assessments, there will most likely be additional, performance-oriented and standard-compliance-oriented requirements and testing needs.

As soon as RIS is used to manipulate the impinging signal with respect to radio parameters, such as beamforming, and signal quality parameters, such as modulation quality, there will be a need to test those capabilities according to the standard. So far, as previously explained, SDOs are still in an early phase regarding RIS specifications, and the potential need for conformance and performance assessment has yet to be specified.

There are various ways to look at RIS deployments: for instance, an RIS can operate as an intelligent reflector to direct the impinging signal to a desired, non-specular reflection angle. Therefore, tests to assess this capability are to be expected. Another perspective could consider the RIS as an intelligent antenna, manipulating the impinging wave with respect to radio beam characteristics or its modulation quality. Consequently, tests assessing the radio beam or signal quality would be required. Regardless of how an RIS is utilized, it seems clear that all conformance and performance tests, as well as all expected regulatory assessments, should be OTA tests. The important question, however, is whether there is a need to specify new, RIS-specific assessment metrics. As the RIS manipulates an impinging signal in a specific way, by default, there might be a need not only to qualify its performance by examining the reflected signal alone but also by considering the comparison with the impinging wave.

\subsubsection{Operational Requirements Gaps}

Possibly the most serious concern of the industry regarding the operation of RIS is the fact that it will be exposed not only to desired signals but also to any other radio signal out there. Thus, co-existence issues need to be addressed with great interest. According to regulations, radio equipment shall not interfere with other services, therefore, co-existence can also be regarded as a regulatory issue. In view of this, coexistence testing scenarios need to be considered, including both in-band (desired signal spectrum) and out-of-band, such as adjacent spectrum signals. Nevertheless, this is a broad area and needs to be addressed by dedicated investigations, which are beyond the scope of this work.

\subsection{Technical Challenges}\label{sec:challenge} 

Despite the considerable potential of RISs in revolutionizing next-generation networks, several challenges need to be addressed, which might prevent successful integration into mobile networks. This section provides a comprehensive discussion of these challenges and refers to recent literature. 

\subsubsection{Channel Estimation and Radio Signal Overhead}
Accurate channel state information (CSI) is vital for the efficient operation of RISs. This information enables the control of passive elements to achieve optimal signal reflection and manipulation. As narrow beams are required to overcome path loss, a high number of RIS elements is needed. This leads to a high control signaling overhead for controlling the RIS. The already challenging overhead is further affected by the following effects: In general, a cascaded channel between the base station, RIS, and UE must be estimated. However, obtaining accurate CSI of this channel is challenging due to potential near-field characteristics, mutual coupling, and beam squint effects, which can lead to errors in channel estimation and performance degradation. Furthermore, a recently published study on optimal RIS configuration in rich scattering environment indicates that the linear cascaded models are not correct~\cite{shen2022,rabault2023}. All named challenges add up, increasing the required overhead to optimal configure the RIS. To address the challenge of a tremendous overhead, passive beam training techniques employing predefined beam pattern codebooks were suggested~\cite{jiao2022,siddiqi2022,okogbaa2022, gong2020,an21,an24no2}. These approaches seek to exploit the inherent structure of the RIS-assisted channel to obtain accurate CSI required for optimal RIS manipulation but misses the ability for flexible deployments. Further research must investigate balancing the trade-off between RIS coverage and resource overhead while maintaining communication quality.

\subsubsection{Regulatory and Multi-Operator Interference Issues}
One of the major challenges for RIS deployment in multi-operator scenarios is managing interference among different network operators, ensuring fairness in resource allocation, and avoiding destructive interference. As RISs can potentially modify the propagation environment across multiple operators, the allocation of resources and the concurrent usage of multiple RISs deployed by different operators become critical concerns. The wideband properties and non-linear saturation effects of RISs introduce challenges related to unwanted interference and regulatory compliance~\cite{jiao2022}. When an RIS is deployed and controlled by one operator, it can cause unpredictable changes in the propagation environment for other operators in the nearby spectrum. Such alterations in the propagation environment can lead to issues such as additional reflections affecting cell coverage, neighbor-cell interference, and may disturb network planning. Furthermore, rapid variations in reflection patterns can disturb instantaneous spatial multiplexing and link adaptation for both reciprocity-based and downlink CSI-based scenarios. Research in this area is in its nascent stages, but preliminary work suggests the need for collaborative efforts among operators. Addressing these challenges necessitates the development of interference mitigation techniques and careful consideration of regulatory guidelines.

Possible solutions for mitigating this issue include the development of standardized protocols for RIS deployment and interference management, the enhancement of cooperation schemes like cooperative scheduling among network operators, and interference-aware resource allocation algorithms that minimize multi-operator interference. Further research must focus on identifying and quantifying the key factors influencing multi-operator interference and exploring advanced techniques for interference mitigation and management in RIS-assisted networks.

\subsubsection{Competitive Technologies}\label{sec:ncr}

To assess the impact of a RIS in a network, a simple model based on~\cite{bjornson2019} is studied. Here the power $P$ of a base station is calculated as a function of the data rate $R$, the channel $\beta$, and the noise power $\sigma$. This is done for the cases where only a base station (BS) and a destination (UE) are considered, compared with the case where the RIS is in the network between the base station and UE, as well as a relay instead of a RIS. The first case is a non-line-of-sight transmission between BS and UE, and the latter two are line-of-sight between the base station and RIS (relay), and RIS (relay) and UE. To describe the channel, the reference channel gain functions for 3 GHz are used, taken from 3GPP for an urban micro scenario. The model of the previously mentioned reference was extended for frequencies between $0.5$ and $100$ GHz taken from~\cite{3gpp22}. Another quantity to assess the performance of the RIS in a network is the energy efficiency, where the most important term is inversely proportional to the power $P$. Implementing the model and using typical parameters (data rate $=4,\text{bit/Hz}$), the base station and the RIS are separated by a distance of $80,\text{m}$. The base station and the destination are separated in a line-of-sight distance of $\sqrt{(10^2 + d_1)}$, where $d_1$ (given in meters) is varied. The RIS size is varied, which is done by the number of RIS elements $N$.

\begin{figure}
    \centering
    \subfloat[6 GHz case.]{\includegraphics[width=0.5\columnwidth]{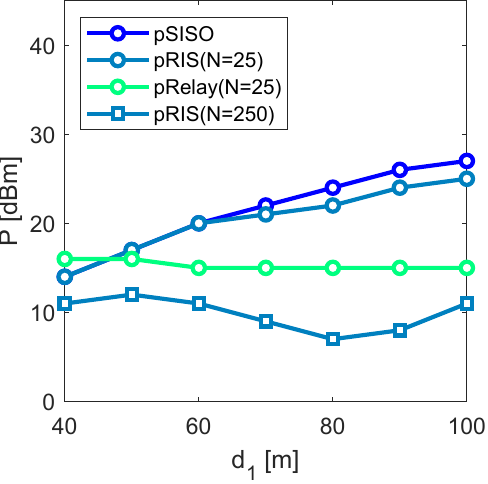}}
    \hfill
    \subfloat[27 GHz case.]{\includegraphics[width=0.5\columnwidth]{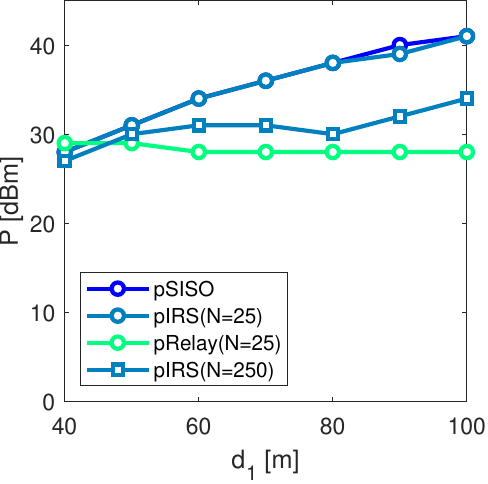}}
    \caption{Transmit power $P$ as a function of the distance $d_1$, which measures the distance between the source and the destination of the signal. The cases of the direct beam (SISO), a relay in between, and the RIS are considered. The RIS can be implemented with an element number $N$ of either $25$ or $250$ tiles per RIS.}
    \label{fig:siso}
\end{figure}

The plot of $P$ for different frequencies is shown in Figure~\ref{fig:siso}. The SISO case shows the highest transmit power $P$ for the considered distance $d_1$, increasing with increasing $d_1$. The relay shows a low and more or less constant value. The RIS case with RIS element number $N = 25$ shows a slight improvement compared to the SISO case. The transmit power with RIS decreases when the RIS element number increases. Furthermore, if the RIS is close to the base station, the transmit power is low. Therefore, in this case, the RIS is competitive or even better than the relay where active gain of the signal is included. If the frequency increases, the power increases. In this case, relatively speaking, the relay needs less power even than the RIS with a higher element number. In summary, RIS is suitable for a position close to the BS (UE), where the advantage diminishes with increasing frequency.

The potential benefits of RISs must also be assessed compared to competitive technologies like decode-and-forward relays~\cite{bjornson2019,carvalho2024network} or network-controlled repeaters (NCRs)~\cite{lin2022overview}. While these solutions may be more costly alternatives compared to RIS, they offer better controllability and potentially larger coverage, as RISs are more susceptible to multiplicative fading. The path loss of the BS -- RIS or NCR -- UE (BS--{RIS/NCR}--UE) link is a product of the individual link path losses. It can be compensated by amplification capabilities in the NCR case. In the case of RIS, it requires a close deployment of the RIS to either the BS or the UE to enable signal improvement or a very large RIS with an extensive element count (see also Figure~\ref{fig:siso}). Besides the production costs, deployment issues, and maintenance costs, which all increase with size, a very large RIS leads to problems described in the earlier paragraph on channel estimation challenges. For a fair comparison between RIS and NCRs, all named advantages and disadvantages must be added up to a total cost of ownership per deployment. To evaluate the economic choice between RIS and NCR, network operators will have to consider the cost for coverage, which is a sum of the deployment costs per area. Because several cheaper RISs are required to cover the same area as one NCR, as stated earlier, the lower RIS deployment cost must level out this factor. The many aspects of the cost for coverage lead to a complex calculation. This evaluation is seen as a crucial aspect of future research. A more extensive comparison of RISs and NCRs is provided in~\cite{aastrom2024ris}.

\section{Hardware Design}\label{sec:hw}

RISs are reflectors usually based on periodic structures that create a phase difference in the wave. The reconfigurability is obtained by modifying the phase difference between the different cells. The project 6G-LICRIS attempts to improve the state-of-the-art of RIS and investigate new hardware concepts. Some of the ideas being implemented within the scope of the project are briefly introduced. The proposed concepts follow a passive design, reflecting the outgoing RF wave and saving power compared to base station antennas, with the goal of enhancing coverage, spectral efficiency, and beam management while improving energy efficiency. Two different RIS models based on different approaches are proposed. The first model is based on resonant structures, while the second model is based on delay lines coupled to a patch antenna at the $28$ GHz band. Both RIS topologies are planned to be tested in indoor scenarios.

\subsection{Multi-resonance RIS}
The first model consists of a multi-resonant RIS integrated into liquid crystal (LC) with wideband behavior and compatibility with ultra-thin layers. Since the switching time of the liquid crystal is directly proportional to the square of the thickness of the liquid crystal layer~\cite{maune18}, it is important to maintain a thin LC layer to achieve fast switching and make the RIS suitable for changeable environments. This requirement may be problematic since very thin LC layers may create a mismatch between the impedance of the resonators and the free-space impedance. In~\cite{perezpalomino12}, it is shown how multi-resonant structures can be used in combination with LC, but a LC layer with a thickness of around $0.05\lambda_0$ is used, and the project aimed at obtaining a working RIS with an LC thickness of around $0.002 \lambda_0$, where $\lambda_0$ represents the free-space wavelength.

\begin{figure}
    \centering
    \subfloat[Stackup of the multi-resonance unit cell proposed.]{\includegraphics[width=\columnwidth]{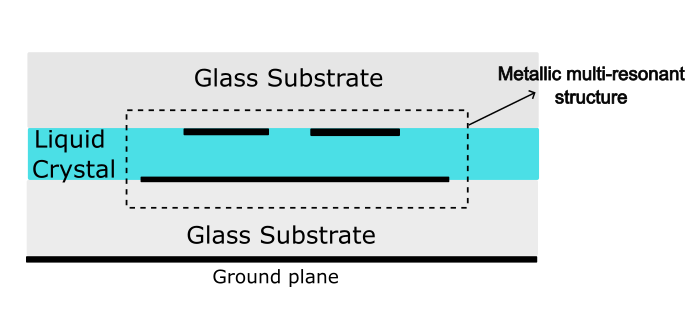}}
    \hfill
    \subfloat[Schematic front view of the RIS, where the top resonant structures are shown.]{\includegraphics[width=0.75\columnwidth]{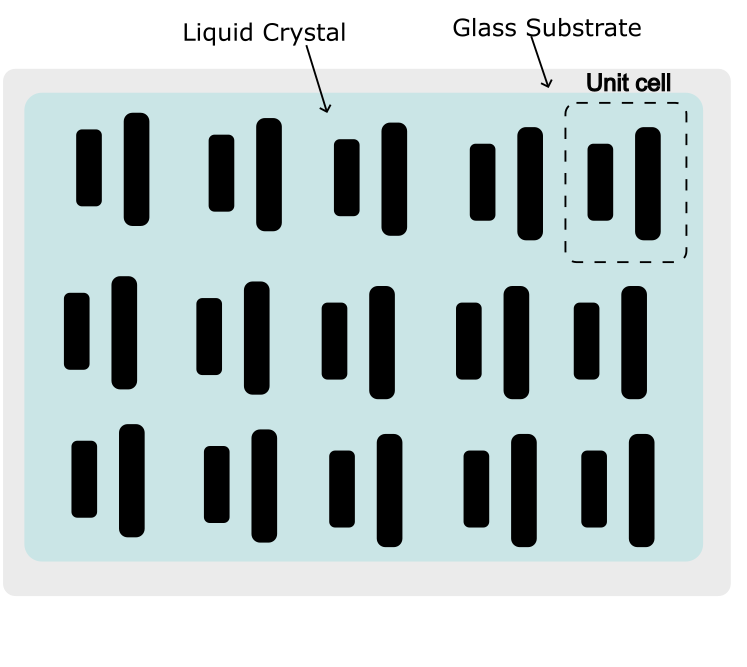}}
    \caption{Multi-resonance RIS design.}
    \label{fig:multiresonantris:design}
\end{figure}

Some attempts were performed to achieve good performance with such a thin layer. For instance, in~\cite{aghabeyki23}, the authors presented a RIS based on resonators compatible with thin LC layers, but the behavior is relatively narrowband. The presented topology consists of a layer of LC sandwiched between two layers of glass substrate. Different metallic resonant structures are printed onto the glass using photolithography processes. Therefore, they are placed directly on top and bottom of the liquid crystal, and their resonances are influenced by the polarization of the LC molecules. When multiple resonances are used, a wide bandwidth is expected. In this case, a bandwidth of around $6$ GHz with a central frequency of $28$ GHz is obtained by simulation. The proposed unit cell is compatible with a single linear polarization. Front and stack-up views of the RIS unit cell are shown in Figure~\ref{fig:multiresonantris:design}.

\begin{figure}
    \centering
    \subfloat[Amplitude of the unit cell vs. frequency for different LC permittivity values.]{\includegraphics[width=\columnwidth]{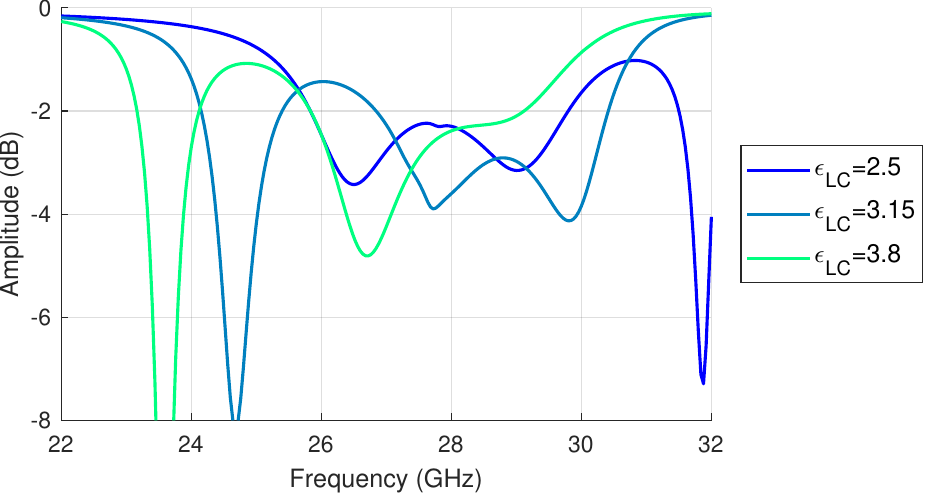}}
    \hfill
    \subfloat[Phase of the unit cell vs frequency for different LC permittivity values.]{\includegraphics[width=\columnwidth]{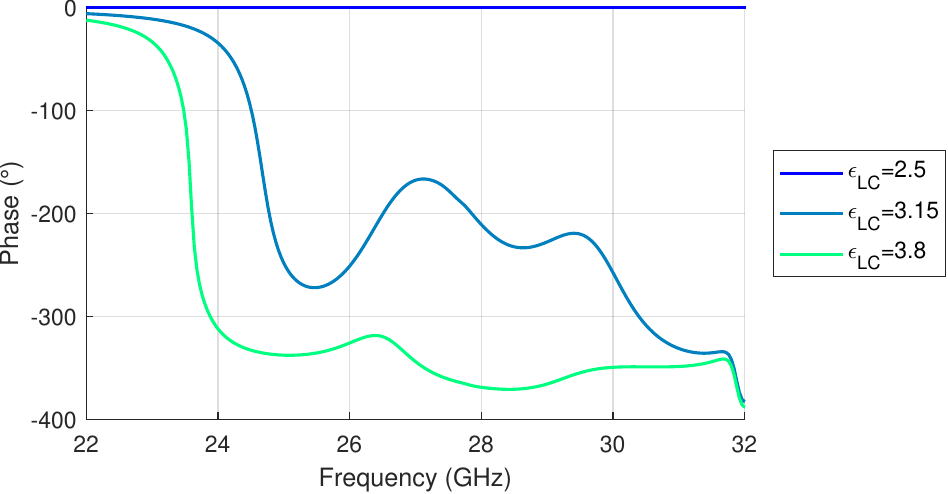}}
    \caption{Multi-resonance RIS simulation results.}
    \label{fig:multiresonantris:results}
\end{figure}

The expected performance of such a RIS was simulated using CST Studio~\cite{cst}, and the amplitude of the reflected wave as well as the phase for a single unit cell is shown in Figure~\ref{fig:multiresonantris:results}, where the phase is normalized to the phase obtained for the minimum permittivity value. It can be observed that a wideband performance with a moderate amount of losses is obtained. The phase difference operates between approximately $24$ and $32$ GHz with a difference of around $320^\circ$ and a constant phase difference across the entire frequency range.

\subsection{Delay-line based RIS}
The second type of RIS consists of a coupled patch RIS. The working principle of this RIS topology is such that when the wave reaches the patch, it is coupled into the active phase shifter through two slots in the ground plane, where each slot works for a different polarization. In this topology, the radiating element is a patch that is coupled to a reflective LC phase shifter. Once the wave passes through the entire phase shifter, it is reflected at the end due to a termination in an open circuit, after which the wave is re-radiated. There have been some attempts to develop such a LC phase shifter-based RIS~\cite{aghabeyki23no2,li21}, but due to space constraints, the lines are printed in a spiral shape, which does not allow for the use of dual polarization. To enable two polarizations, it is necessary to have the space to place two perpendicular phase shifters. Therefore, the lines should be miniaturized. The proposed method to achieve this is to design the phase shifter using miniaturization techniques, thereby increasing the phase shift per unit of length. The phase shifter may provide a variable phase shift ranging from $0^\circ$ to $360^\circ$. A bandwidth of around $3$ GHz is expected, and it is compatible with two orthogonal polarizations.

\begin{figure}
    \centering
    \subfloat[Stackup view of the phaseshifter delay RIS.]{\includegraphics[width=0.75\columnwidth]{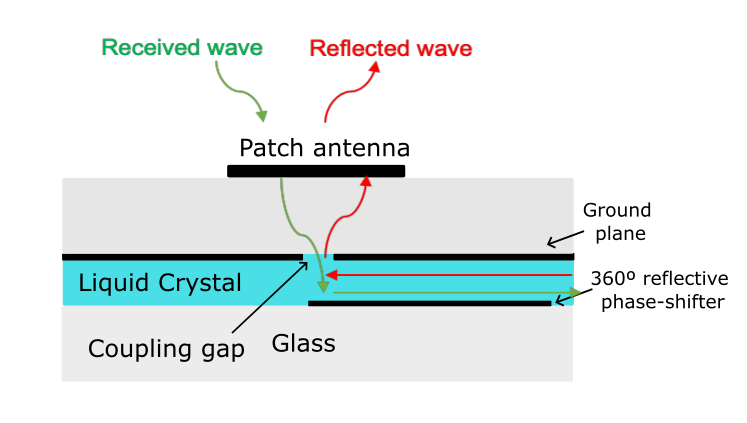}}
    \hfill
    \subfloat[Top view of the unit cell based on a patch antenna over a glass substrate and a ground plane with two slots.]{\includegraphics[width=0.75\columnwidth]{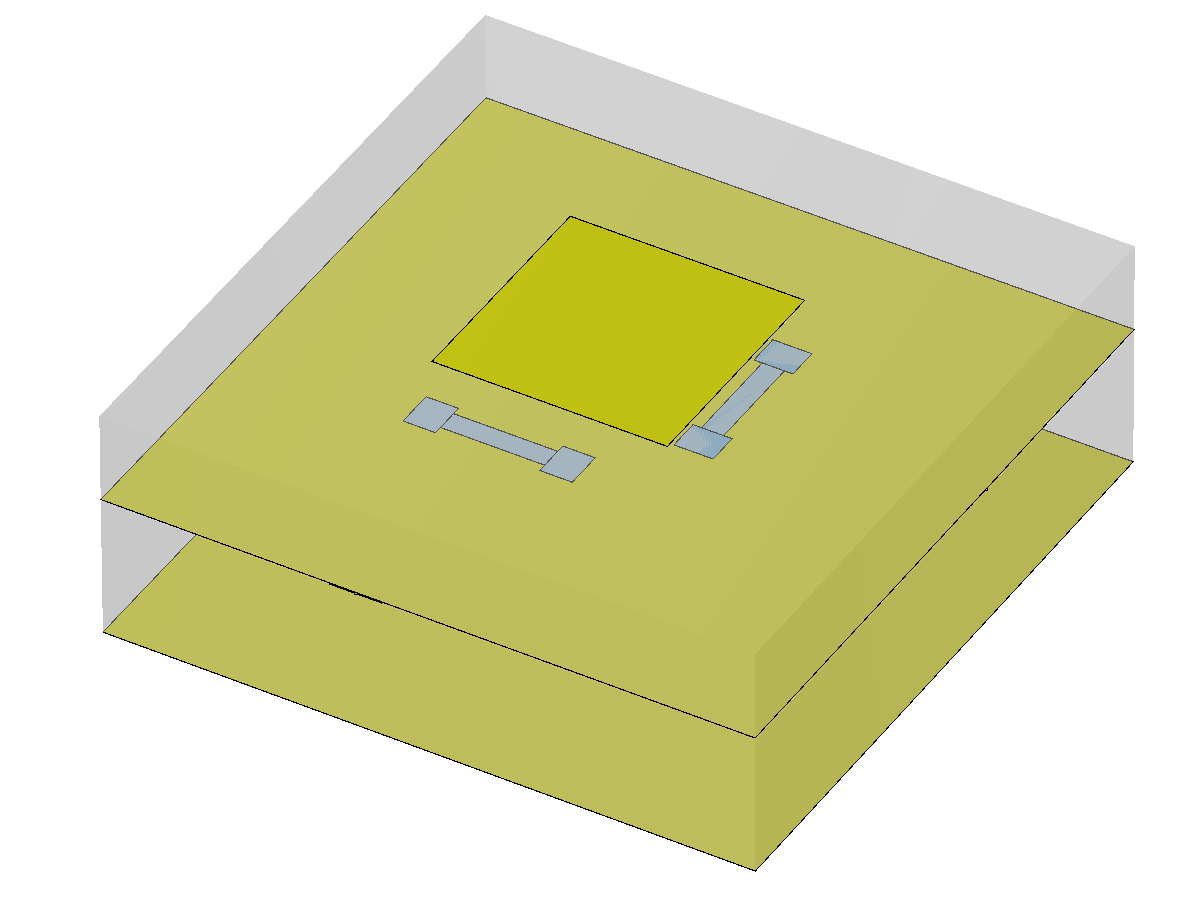}}
    \hfill
    \subfloat[View of the slots in combination with the LC-phase shifters topologies.]{\includegraphics[width=0.75\columnwidth]{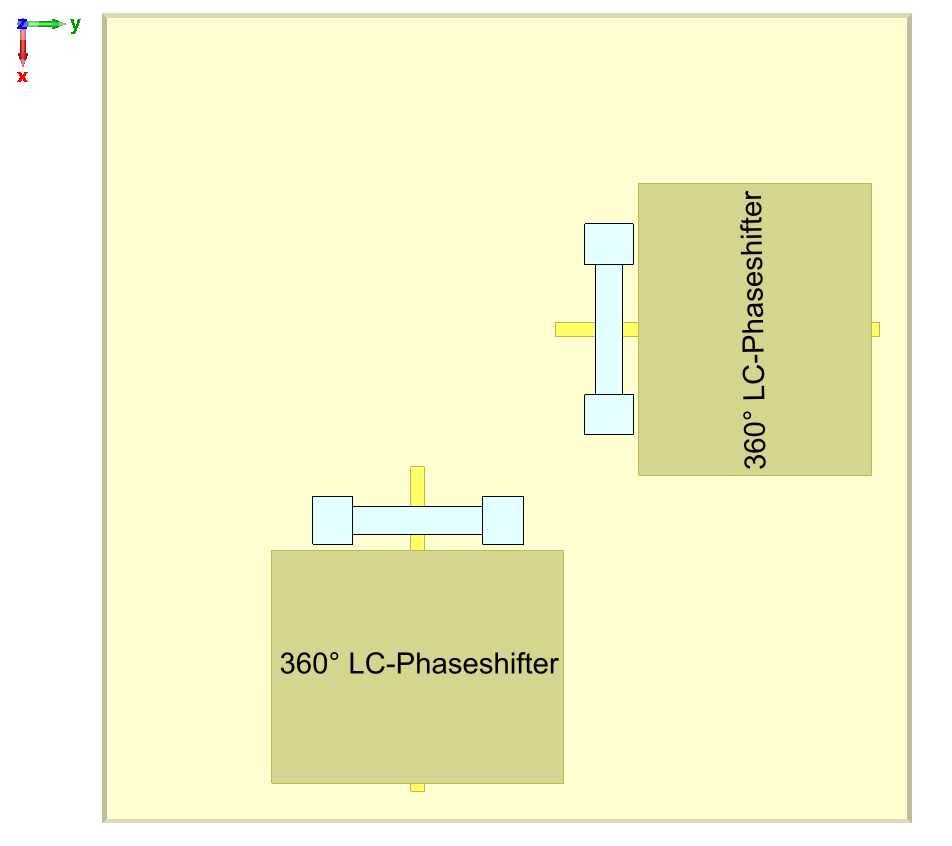}}
    \caption{Delay-line based RIS design.}
    \label{fig:psris:design}
\end{figure}

\begin{figure}
    \centering
    \subfloat[Amplitude of the reflected wave at the unit cell for one polarization.]{\includegraphics[width=\columnwidth]{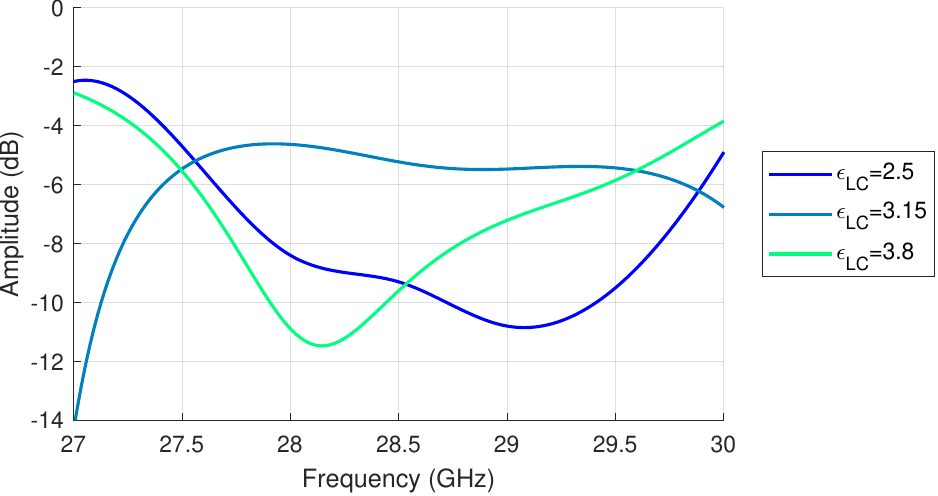}}
    \hfill
    \subfloat[Phase of the reflected wave at the unit cell for one polarization.]{\includegraphics[width=\columnwidth]{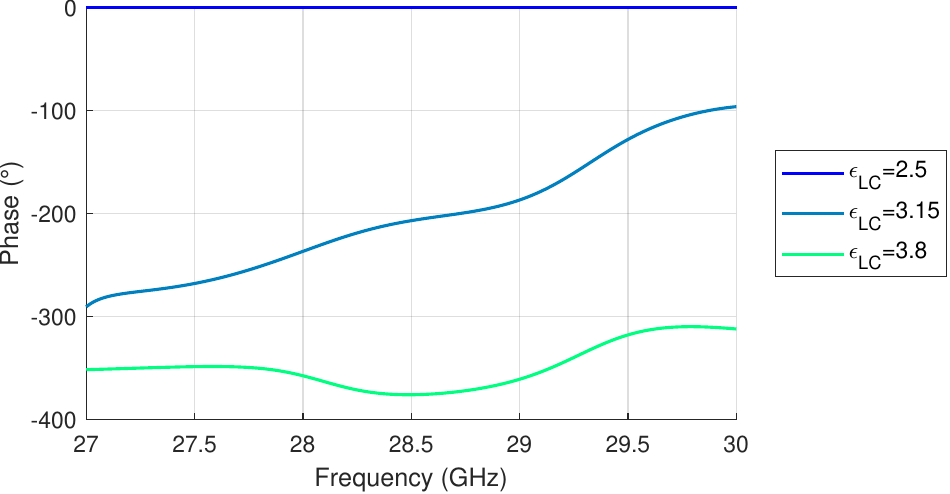}}
    \caption{Delay-line based RIS simulation results.}
    \label{fig:psris:results}
\end{figure}

A preliminary view of the stack and a perspective view that allows seeing the patch over the two slots, which are coupled to the phase shifter, are shown in Figure~\ref{fig:psris:design}. The performance of the unit cell can be seen in Figure~\ref{fig:psris:results}. The amplitude and phase of the unit cell for different values of permittivity are shown, where the phase is normalized to the phase for the minimum permittivity value.


\section{Deployment and Dimensioning Aspects}\label{sec:depl}

This section discusses the challenges related to RIS deployment and dimensioning. One of the main challenges in RIS deployment is related to the calculation of RIS placement in a cellular network, while RIS dimensioning involves calculating the number of reflecting elements necessary to achieve the required system targets. Central to deployment and dimensioning is the propagation model considered in the calculations. To this end, this section begins with a discussion on RIS propagation modeling, followed by deployment optimization, and RIS dimensioning.

\subsection{Propagation Modeling} \label{sec:propmodel}

In order to manage and control RISs from a system-level perspective, it is necessary to understand and ideally anticipate the effects of their configuration on intended changes in radio propagation as well as unintended effects, e.g., in-band and out-of-band interference. This requires fast and reliable RF propagation models, where a reasonable loss of accuracy may be acceptable.

\begin{figure}
    \centering
    \includegraphics[width=\columnwidth]{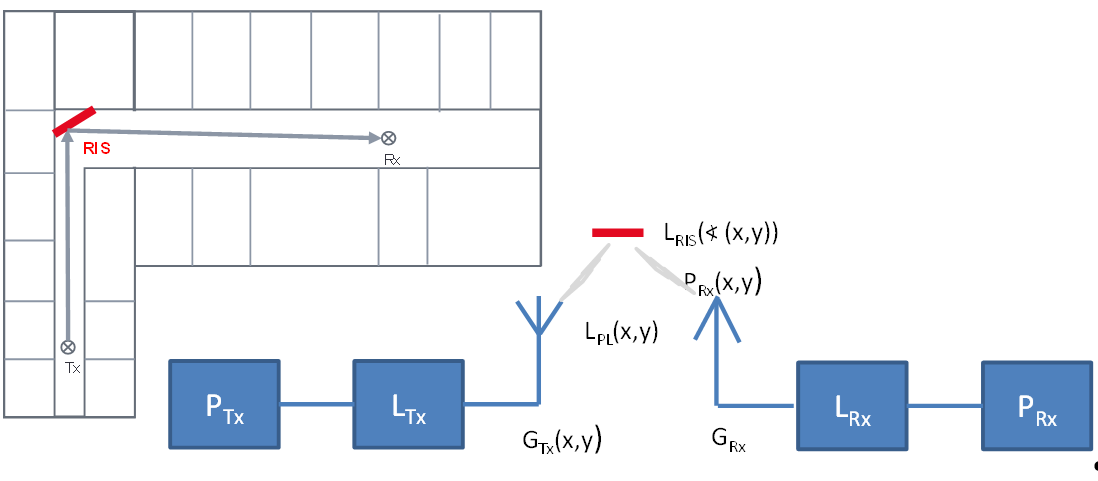}
    \caption{Path model of RF propagation components including an RIS working as a ``directed mirror''. The blocks are named with ``L'' representing losses, ``G'' representing gains, and ``P'' representing power values. ``Tx'' stands for transmitting, and ``Rx'' for receiving. The loss of the RIS depends on the angle of arrival/transmission from the sender/receiver.}\label{fig:mod:prop}
\end{figure}

Semi-empirical and/or geometric RF propagation models offer a good compromise. The measures that need to be taken into account are, first of all, the received power (see Figure~\ref{fig:mod:prop}). In a more sophisticated model, which is beyond the scope of the current work, the impact of the MIMO path through the RIS would also be of interest. Since RIS can be most efficiently used at higher frequencies beyond $6$~GHz, RF propagation in scenarios where RISs are used is typically limited by LOS between the transmitter and receiver. The intention of using RIS is to supplement the LOS paths with reflections off the RIS. While semi-empirical models are widely used in sub-$6$~GHz mobile radio scenarios (or historically sub-$2$~GHz), geometric models are less limited in their frequency application ranges, since they are based on deterministic approaches.

Semi-empirical models offer the advantage of relatively low computational complexity compared to other RF propagation modeling approaches, such as ray tracing and full-wave simulations, which can be computationally intensive and time-consuming. However, they also have limitations. These models are based on empirical measurements that may not always capture the full complexity of real-world environments, and they may not accurately predict the effects of new technologies and deployment scenarios that differ from the assumptions of the original model. Therefore, careful validation and calibration of semi-empirical models using real-world measurements and data are essential to ensure their accuracy and reliability.

\begin{figure}
    \centering
    \includegraphics[width=\columnwidth]{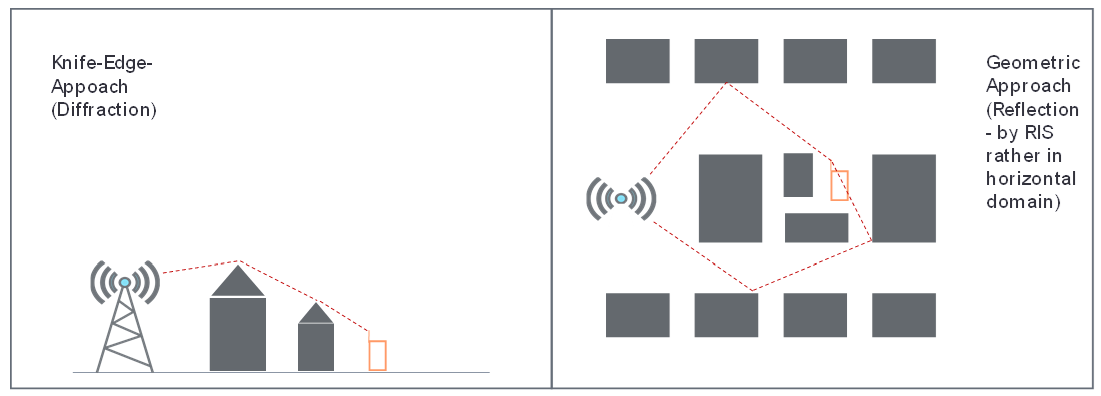}
    \caption{Modeling of diffraction (left) and reflection (right) in RF propagation models.}\label{fig:mod:knife}
\end{figure}

Geometric models are based on the concept that RF signals propagate in straight lines, and their behavior can be predicted by considering the geometry of the environment in which they propagate (see Figure~\ref{fig:mod:knife}, right-hand side). These models use parameters such as the distance between the transmitter and receiver, the height of the antennas, and the surrounding terrain to calculate the path loss of RF signals. For example, the two-beam ground reflection model accounts for reflections of RF signals from the ground by assuming that the signal travels along two paths: a direct path from the transmitter to the receiver and another path reflected from the ground to the receiver. The model considers the heights of the antennas and the distance between the transmitter and receiver to calculate the path loss. The two-beam ground reflection model is commonly used to predict RF propagation in urban environments where ground reflections can significantly affect path loss. A similar approach can be considered for RF predictions with RIS, assuming that the reflection point is from the RIS rather than the ground.

\begin{figure}
    \centering
    \includegraphics[width=\columnwidth]{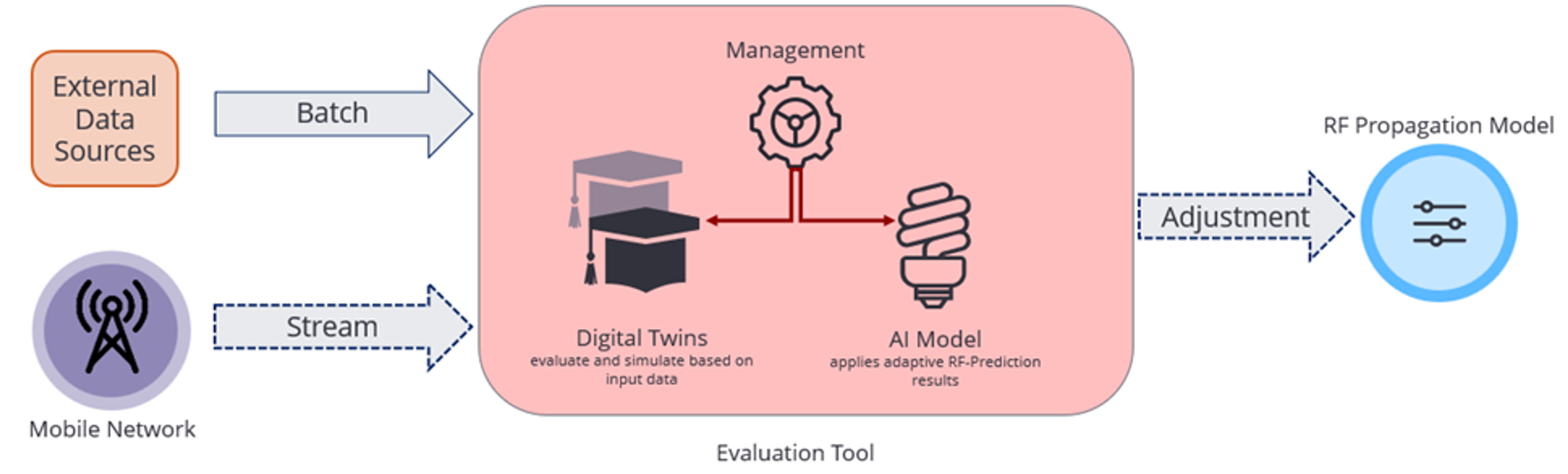}
    \caption{Control Flow for RIS-Network using measurement data and RF prediction.}\label{fig:mod:control}
\end{figure}

More advanced geometric models can also consider additional factors such as the height and location of buildings, the density of obstructions, and the surrounding terrain. The Extended Hata Model~\cite{drocella20163} combines geometric principles with empirical data to account for the effects of both urban and suburban environments, including factors such as building density and road width. It can also account for the effects of obstacles, as is typically done in knife-edge models.

Knife-edge models are named for the sharp edge of the obstacle, which is considered the most significant factor affecting the diffraction of RF signals (see Figure~\ref{fig:mod:knife}, left-hand side). The shape, height, and distance of the obstacle from the transmitter and receiver are key parameters in these models. One of the most commonly used knife-edge models is the single knife-edge diffraction model, which assumes that the RF signal encounters a single sharp edge along its propagation path. The model calculates the path loss based on the height and distance of the obstacle, the height of the antennas, the frequency of the RF signal, and the wavelength of the signal.

For the integration of RIS into RF propagation models, diffraction is certainly of less interest than reflection. However, both modeling approaches could be combined through a combined parameter estimation of their parameters using methods of mathematical optimization, including machine learning/artificial intelligence approaches (see Figure~\ref{fig:mod:control}). Such an approach strongly depends on the availability of suitable training data. For RIS, such data are hardly available. A combination of simulation and ray tracing may provide artificial training data, similar to the approach used by RadioUNet~\cite{levie2021radiounet}.

In any case, the RF characteristics of RIS need to be well understood for different bands, in various environments, and also under different environmental conditions. The proposed RF models need to be flexible in this respect, which outlines the need for simplified models. For control approaches, RF measurement data from the field must constantly feed the control loops, and the RF modeling on the RIS must also be adapted during operations. Therefore, an extensive control loop and measurement collection approach from connected devices within the network is required.

\subsection{Deployment Optimization}\label{sec:deploptim}

Developing an RIS-assisted wireless communication network raises the problem of RIS placement, a part of the so-called network planning problem. In order to realize RIS placement, a model that captures the main propagation characteristics of RIS is needed, while still being mathematically tractable given the large scale of network planning problems. The question of ``what level of abstraction'' is then raised in order to maintain the right balance between accuracy and complexity.

Migration to higher frequency bands such as millimeter wave (mmWave) or the sub-terahertz (sub-THz) to achieve the ambitious requirements in future wireless networks on the one hand, and the product-distance path loss model of RIS on the other hand, imply the need for large surfaces with many elements, which can achieve passive beamforming gain through joint reflection. To compensate for the severe path loss, besides the required large surface, the existence of the LOS path is indispensable, which is equivalent to the LOS link between BS--RIS and RIS--UE. In addition to improving the channel gain, the LOS link also simplifies the control and operation of an RIS, which is a typical assumption in the literature. This issue, moreover, explains a prospective use case of RISs, namely coverage extension, by providing virtual LOS links through the RISs.

Furthermore, another important feature that is largely neglected in the literature is the orientation of the surface. More specifically, the orientation of the surface with respect to the BS and UE. It is known that the further one moves away from the surface broadside (i.e., toward the end fire), the higher the effects of impairments such as beam squint and gain loss become. Therefore, in a realistic model, the field of view (FoV) of the surface needs to be limited, which is determined, for instance, based on the power budget and ensures that the BS and considered area of interest (i.e., technically where UEs are supposed to be) are within this FoV.

\begin{figure}
    \centering
    \includegraphics[width=0.75\columnwidth]{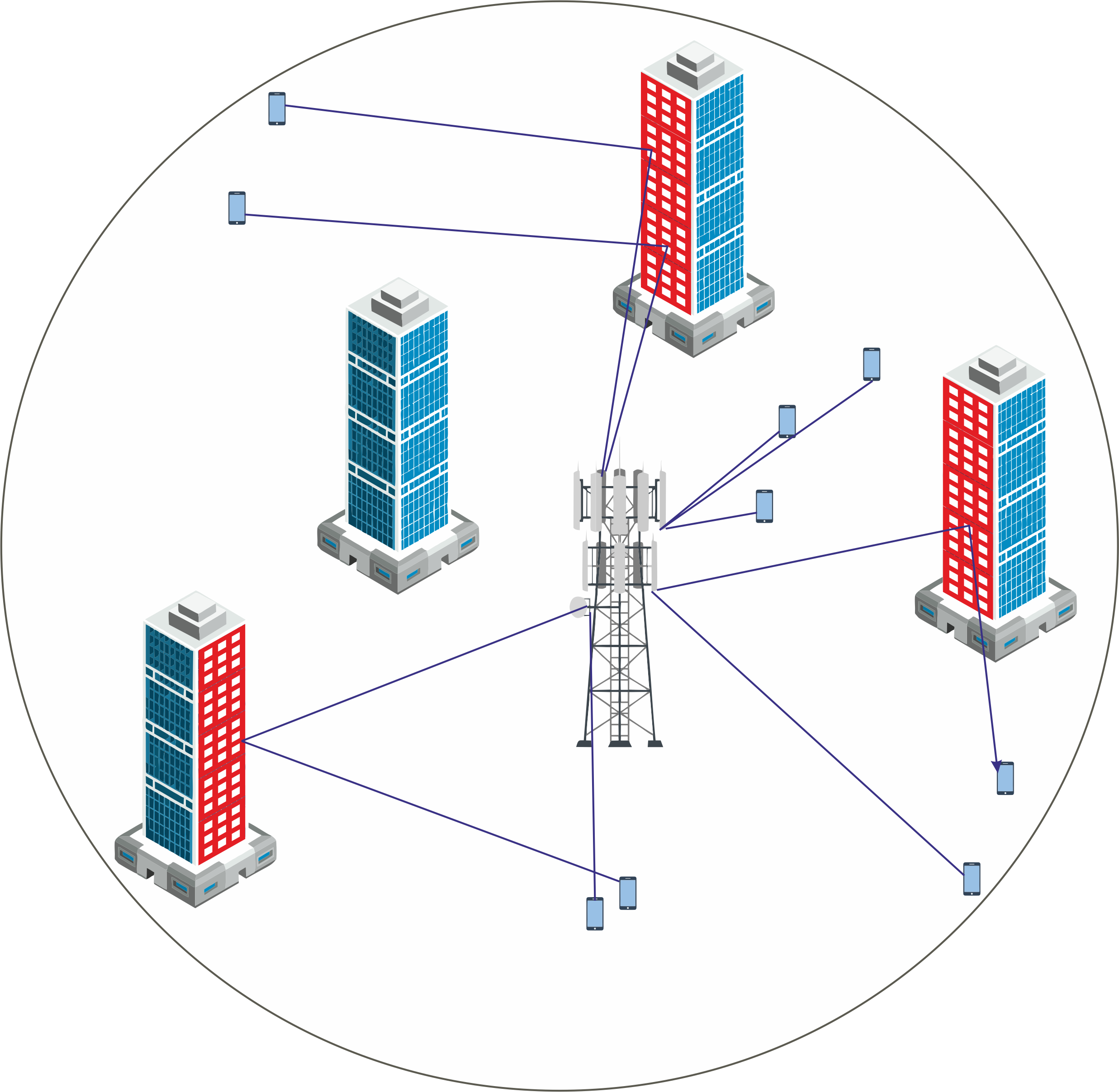}
    \caption{Scenario of a multi-RIS-assisted wireless communication system.}\label{fig:hhi:deployed}
\end{figure}

A multi-RIS-assisted wireless communication system is considered in a scenario as shown in Figure~\ref{fig:hhi:deployed}, which comprises a BS located at $\textbf{P}_B = [b_x, b_y, b_z]^T$, an RIS deployment area $P= \{ P_1, \ldots, P_M \}$ with $M$ RIS candidate locations, and an area of interest $L= \{ L_1, \ldots, L_N \}$  with $N$ potential locations for UEs, to be covered either by the BS or an RIS. A LOS link to the BS location, $\textbf{P}_B$, for all $P \in \mathcal{P}$ is assumed. Moreover, regarding the orientation of the surface, the valid FoV of the surface is defined as  $[-\theta_{\max}, \theta_{\max}] $ with $\theta_{\max}$ being the maximum acceptable deviation from the broadside (for both incident and reflecting angles), that satisfies the given system constraints (e.g., requirements on the maximum beam squint effect and gain loss). 

\begin{figure}
    \centering
    \includegraphics[width=0.75\columnwidth]{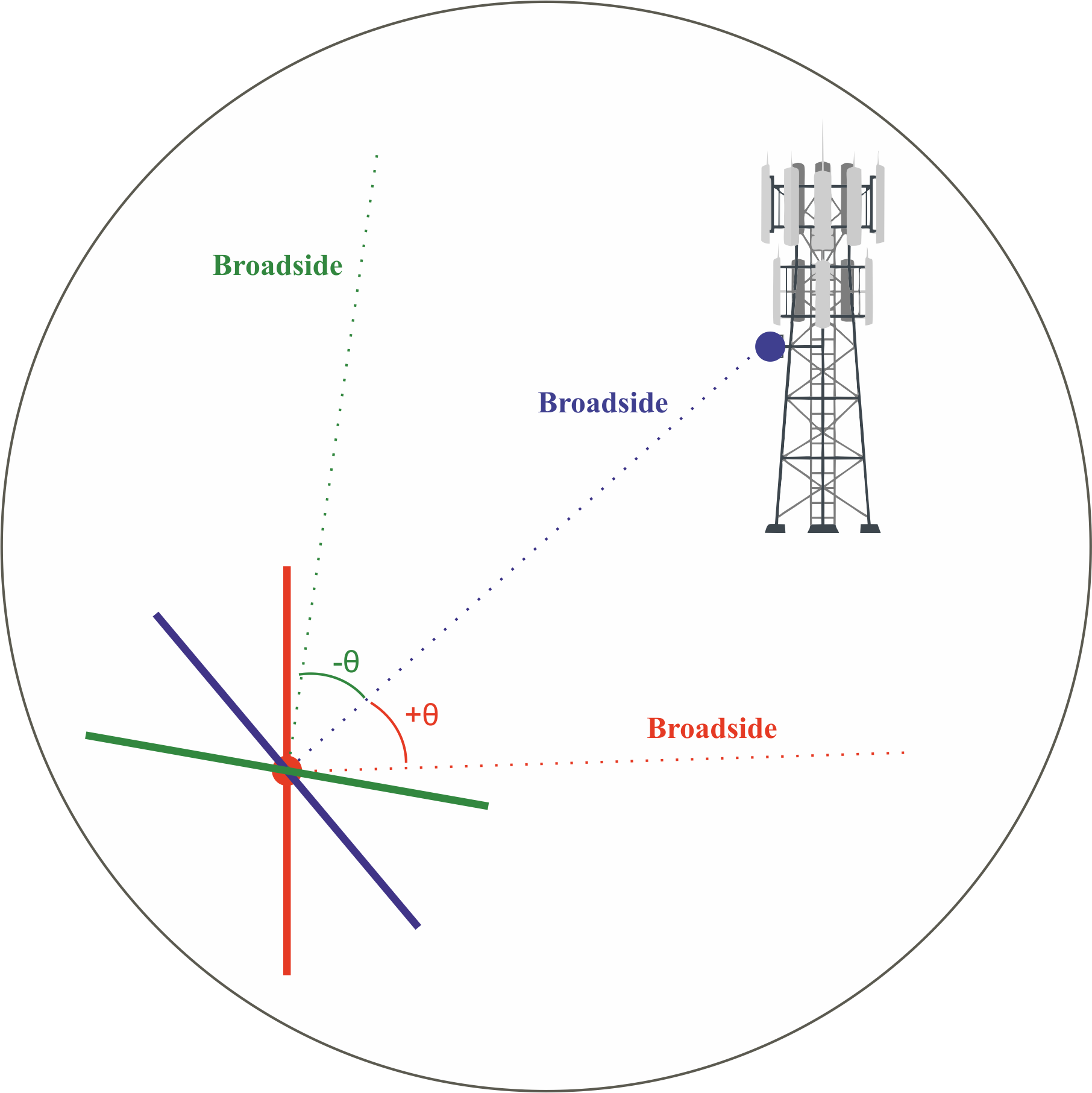}
    \caption{Example of a valid range of azimuth angles for an arbitrary candidate position of RIS and the given BS location.}\label{fig:hhi:range}
\end{figure}

For each candidate location of the RIS, the range of valid orientations of the surface is determined in a way to ensure the steering angle towards the BS is in the range of valid FoVs. Figure~\ref{fig:hhi:range} illustrates an example of a valid range of azimuth angles for an arbitrary candidate position of RIS and the given BS location. The blue surface (line in the figure) is the one with broadside in the direction of the BS, and the red and green ones are the ones that reach $\pm \theta_{\max}$  in steering angle toward the BS. Therefore, the valid range of orientations of the surface that provide a steering angle within the valid FoV are all orientations between the orientation of red and green surfaces. Similarly, for each candidate location of the RIS and for a given UE position, the range of valid orientations of the surface is calculated. This procedure should be repeated for all UE positions that are supposed to be covered by the considered RIS location (i.e., a subset of $L$). As a result, for each RIS candidate location, given the BS location, and the set of UE positions that are supposed to be covered by the considered RIS, the final valid orientations of the surface are calculated by the intersection of all individually determined valid orientations of the surface, i.e., given by BS and the subset UE positions.


\begin{figure}
    \centering
    \includegraphics[width=\columnwidth]{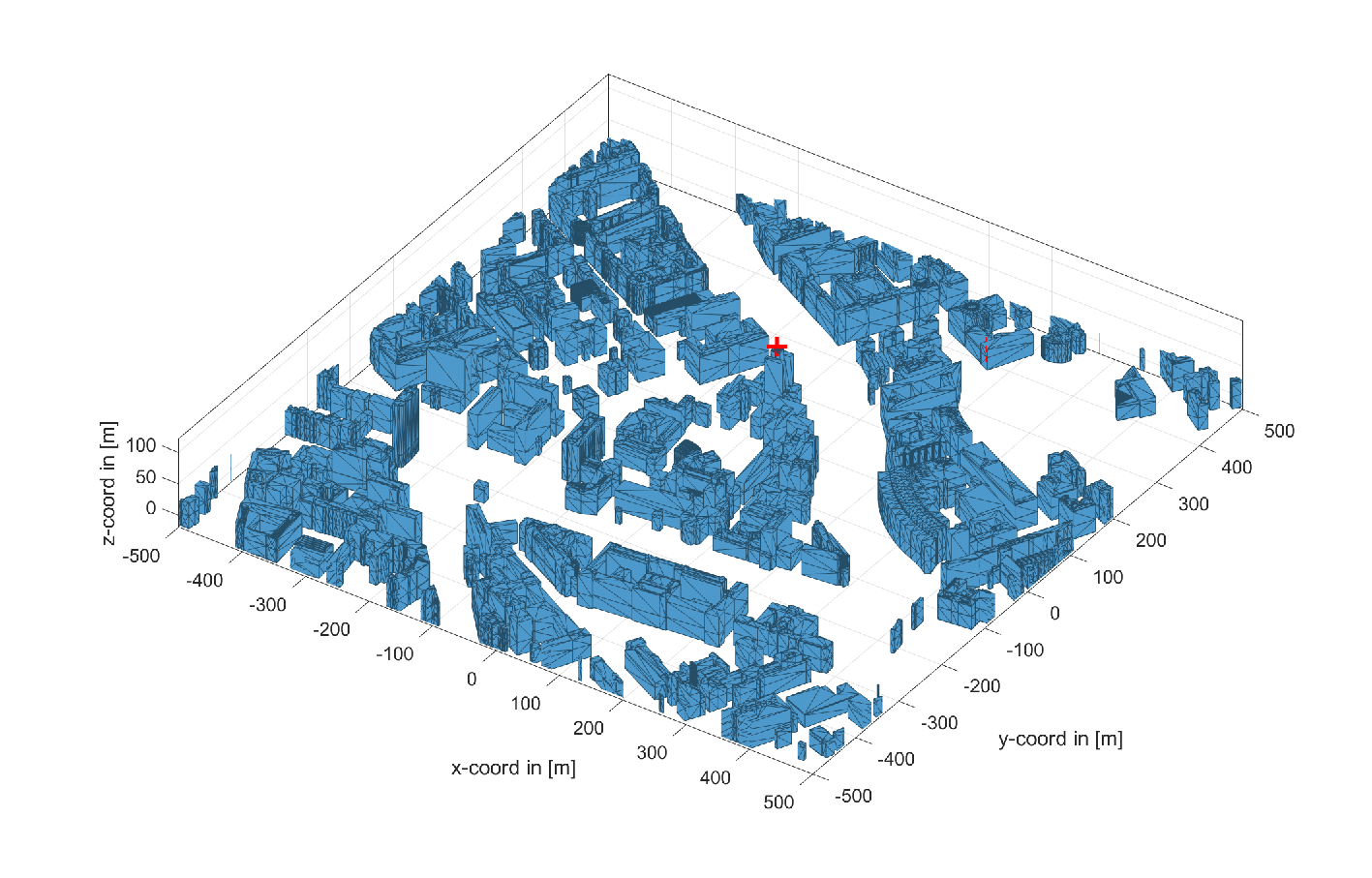}
    \caption{Environment model of the 5G Berlin Testbed.}\label{fig:hhi:testbed}
\end{figure}

In the following, a toy example of extended coverage provided by RISs is illustrated, demonstrating the importance of placement and orientation of the surfaces. For the purpose of the toy example, the channel model ``Quasi Deterministic Radio Channel Generator'' QuaDRiGa~\cite{jaeckel2014quadriga, tohidi2023near} developed at Fraunhofer Institute for Telecommunications, Heinrich-Hertz-Institut (HHI) is used, which has recently been extended by the feature of LOS-detection. This allows site-specific simulations by incorporating data from the environment. Here, the data includes 3D information of buildings and height profiles of the terrain under consideration, which is the campus network ``Berlin 5G testbed'', shown in  Figure~\ref{fig:hhi:testbed}. The location of the BS, $P_B$, is on top of the HHI building in Berlin, Germany.

\begin{figure}
    \centering
    \includegraphics[width=0.75\columnwidth]{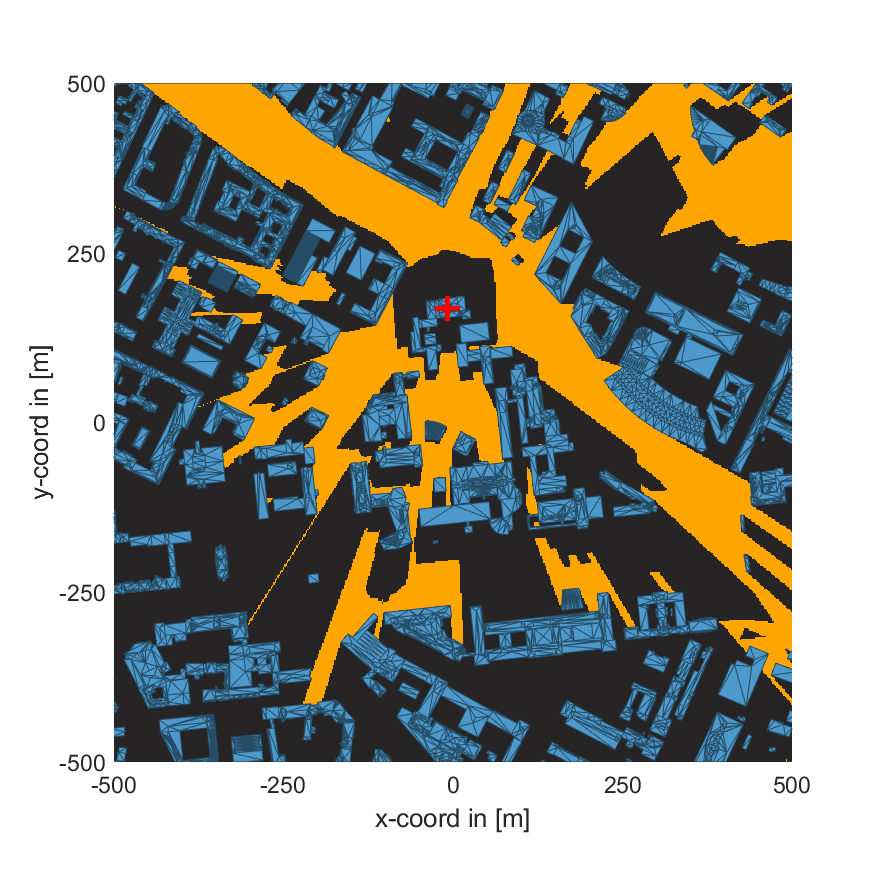}
    \caption{LOS coverage for the direct link from BS to UEs.}\label{fig:hhi:los}
\end{figure}

In Figure~\ref{fig:hhi:los}, the LOS coverage for the direct link from the BS to the UEs is shown, i.e., only UEs at orange positions have an LOS path to the BS. The aim is therefore to expand the coverage area by smart positioning and orientation of RISs and to obtain increased coverage through virtual LOS paths.
 
\begin{figure*}[!t]
  \centering
  \includegraphics[width=6.5in]{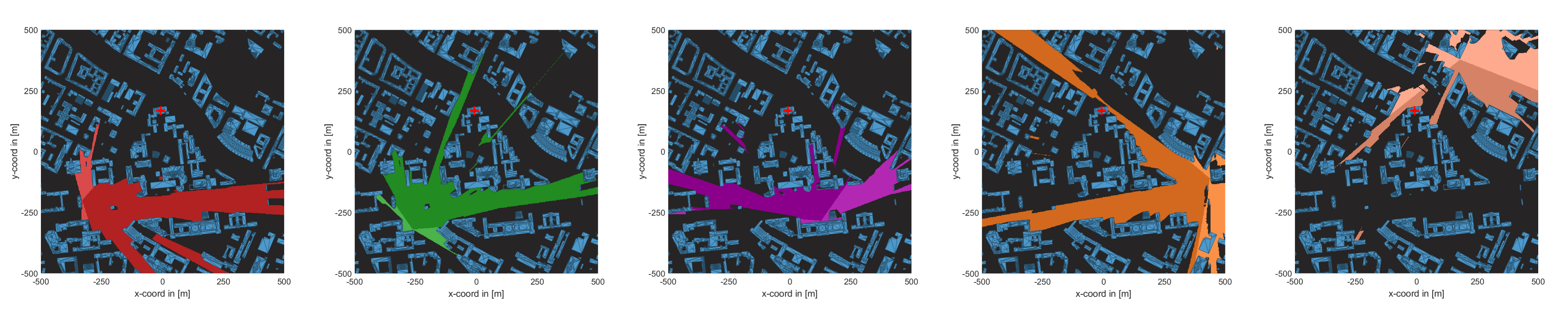}
  \caption{LOS coverage of RIS candidate locations.}
  \label{fig:hhi:candidate}
\end{figure*}

\begin{figure}
    \centering
    \subfloat[Single RIS deployment]{\includegraphics[width=0.75\columnwidth]{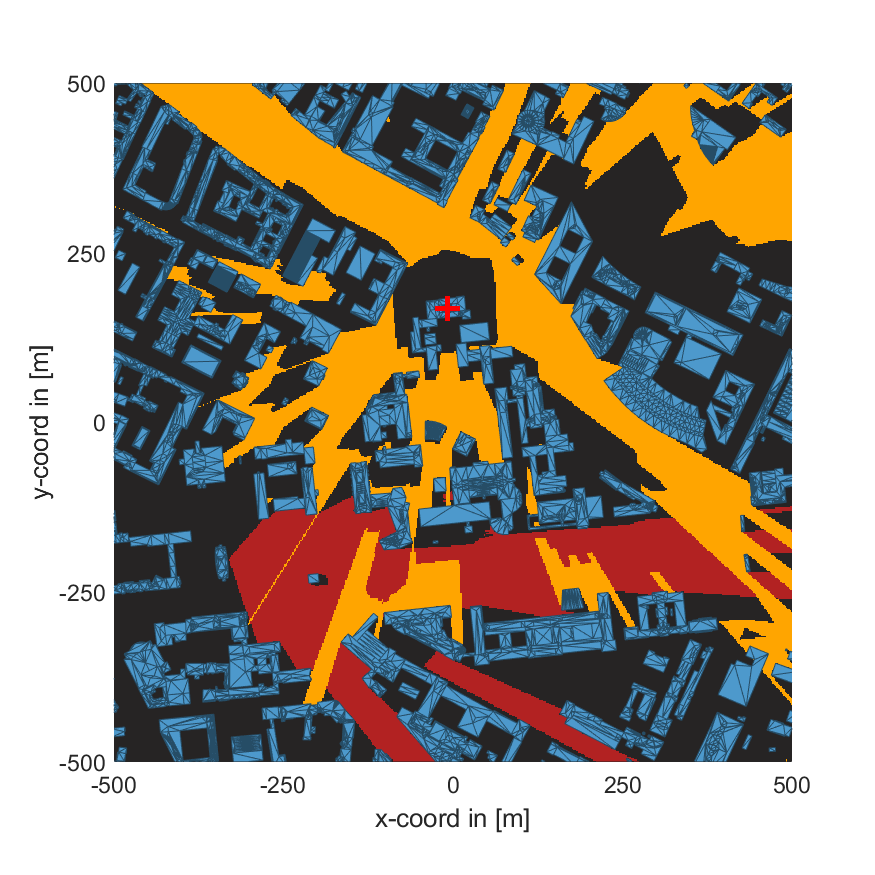}}
    \hfill
    \subfloat[Dual RIS deployment]{\includegraphics[width=0.75\columnwidth]{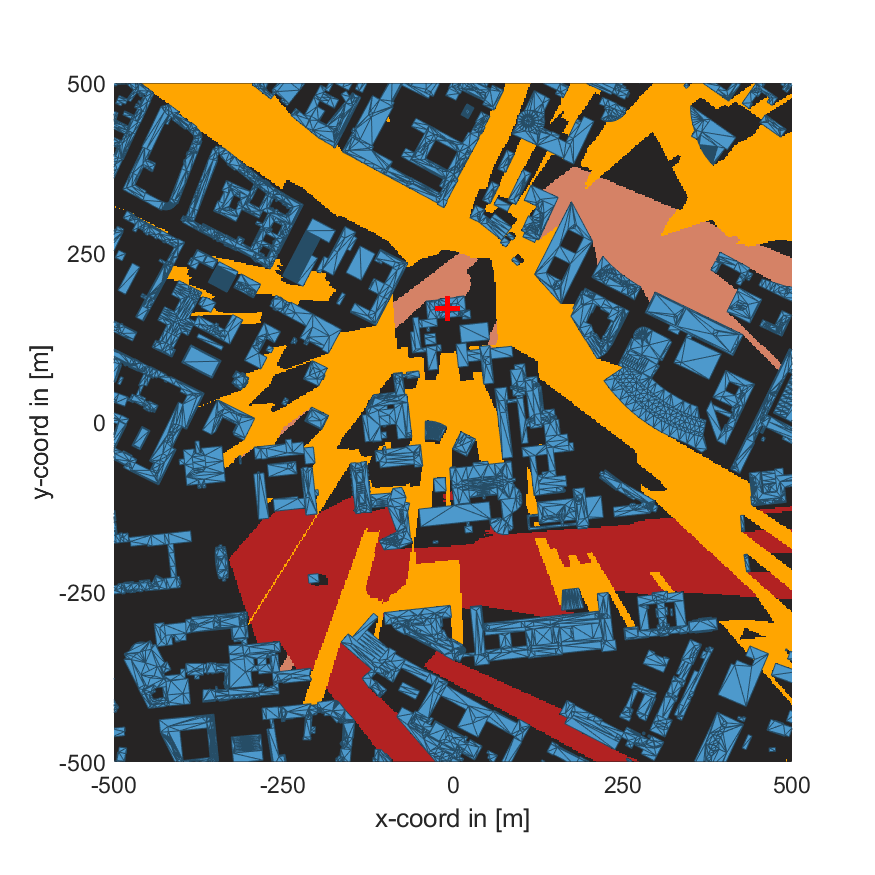}}
    \caption{Extended LOS coverage by BS and RIS(s).}
    \label{fig:hhi:extcov}
\end{figure}

Here, a small-scale scenario is considered in which a set $\mathcal{P} = { P_1, \ldots, P_5 }$ of $5$ RIS candidate positions, each at a height of $30$ meters, is available. Their locations and respective coverage are shown in Figure~\ref{fig:hhi:candidate}. White bars represent the RIS (not to the right scale with respect to the buildings) and the respective total colored areas indicate the LOS coverage by the corresponding RIS without taking into account the RIS orientation. Darker colored areas show the LOS coverage with the RIS orientation that maximizes LOS coverage, and dashed white lines correspond to the respective broadside of the deployed RIS. The deployment for the optimal RIS, i.e., the one that extends the coverage the most, is shown in Figure~\ref{fig:hhi:extcov}. In the bottom part of this figure, the second RIS is deployed, which optimally (in a greedy sense) extends the superposition of the LOS coverage areas of the BS and the first RIS.
 
\begin{figure}
    \centering
    \includegraphics[width=\columnwidth]{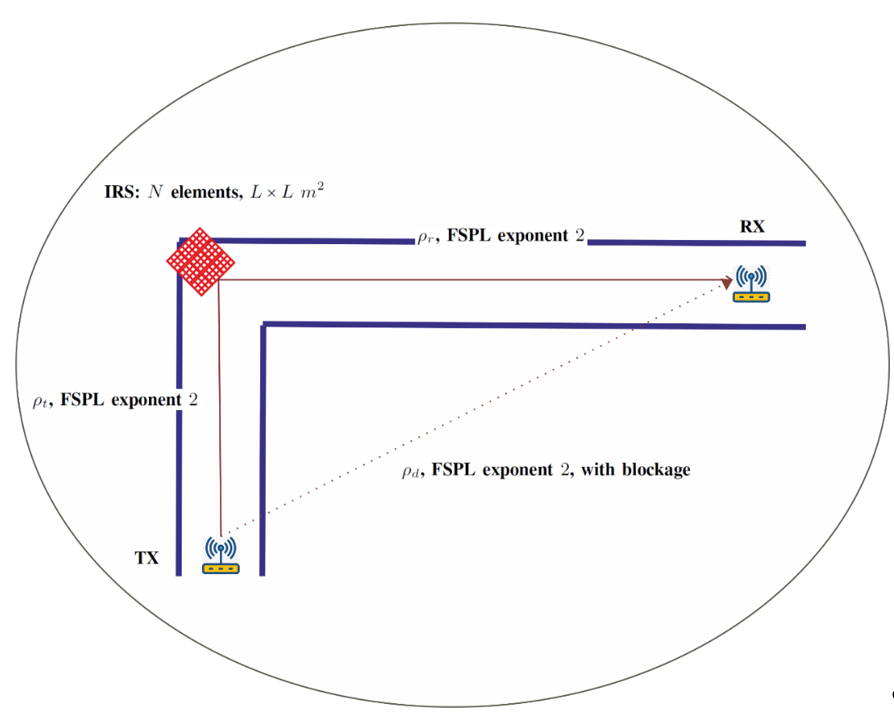}
    \caption{Scenario of an RIS-aided network for the consideration of path loss modeling.}\label{fig:hhi:pl}
\end{figure}

\subsection{Dimensioning}\label{sec:dimensioning}

An efficient deployment strategy for RISs in mobile networks includes not only the selection of suitable locations and orientations for the RISs but also their dimensioning, because the number of elements and the size of the RISs significantly determine the achievable reflection coefficient. Initial modeling of path loss can provide information for dimensioning (and be extended in more comprehensive link budget analyses). For this purpose, consider the scenario of an RIS-supported network from Figure~\ref{fig:hhi:pl}, which includes an RIS with $N$ elements and size $L \times L\, \text{m}^2$. The distances of the three paths $\rho_d$, $\rho_t$, and $\rho_r$, which correspond to the path between the transmitter and receiver, the transmitter and RIS, and the RIS and receiver, respectively, each have a free-space path loss (FSPL) exponent of 2. In addition, the direct path between the transmitter and receiver is blocked, so the signal experiences attenuation at a certain level.

\begin{figure}
    \centering
    \subfloat[Number of elements for $\rho_t=20\,\text{m}$]{\includegraphics[width=\columnwidth]{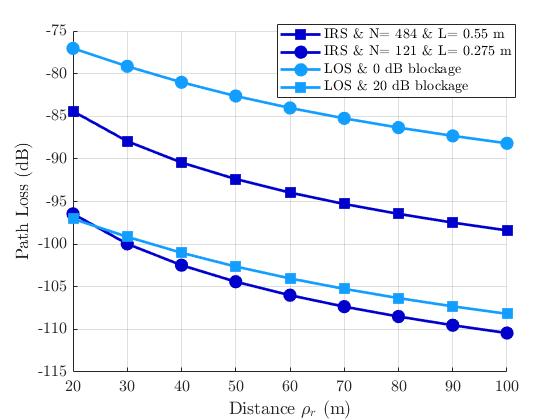}}
    \hfill
    \subfloat[RIS length for $\rho_t=\rho_r $]{\includegraphics[width=\columnwidth]{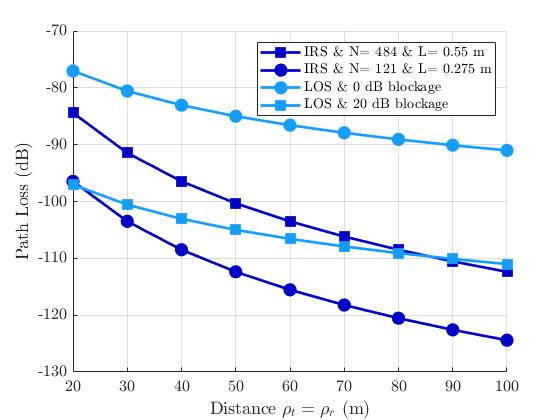}}
    \caption{Path loss against distance at $6$ GHz.}
    \label{fig:hhi:f8}
\end{figure}

In Figure~\ref{fig:hhi:f8}, the resulting path losses are plotted against the distance $\rho_r$ for $\rho_t=20\,\text{m}$ in (a) and for $\rho_t=\rho_r$ in (b) at a frequency of $6$ GHz. While an RIS with $N = 121$ and $L=0.275\,\text{m}$ is not sufficient to compensate for the path loss of the direct path with a $20$ dB blockage, this is possible with an RIS with $N = 484$ and $L=0.55\,\text{m}$. However, this RIS configuration also lags behind a direct LOS link without blockage as well as behind a LOS link with $20$ dB blockage at larger distances $\rho_t = \rho_r \geq 85\, \text{m}$.
 
According to~\cite{najafi2020}, at least an RIS area of
\begin{equation}
    A_\text{req} = L_\text{req} \times L_\text{req} = \frac{\lambda \rho_t \rho_r}{ \rho_d}
\end{equation}
or a number 
\begin{equation}
    N_\text{req} = \frac{4 \rho_t \rho_r}{\lambda \rho_d}
\end{equation}
of elements, with an assumed element size of $\lambda/2$, is required so that the free-space path loss of the RIS-assisted link is equal to that of the LOS link without blockage.

\begin{figure}
    \centering
    \subfloat[Number of elements]{\includegraphics[width=\columnwidth]{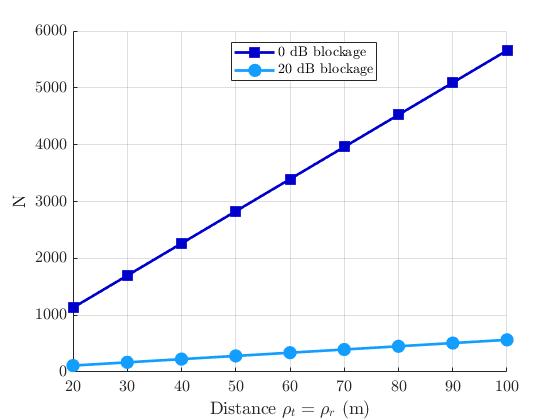}}
    \hfill
    \subfloat[RIS length]{\includegraphics[width=\columnwidth]{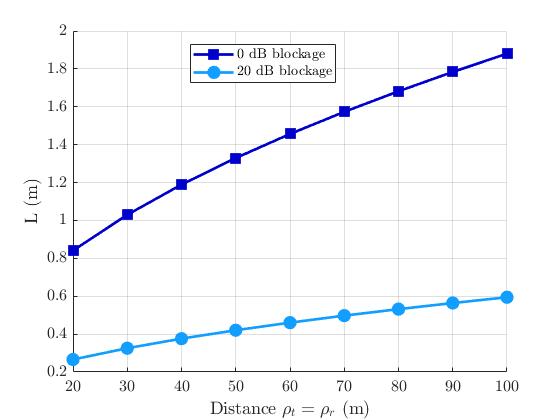}}
    \caption{Required number of RIS elements and required RIS size against distance at $6$ GHz.}
    \label{fig:hhi:f9}
\end{figure}

Based on this, the required number of elements and required RIS size are shown in Figure~\ref{fig:hhi:f9} (over distance $\rho_r =\rho_t$ and for a frequency of $6$ GHz). It can be seen that compensation for path loss, especially for the non-blocked LOS path or high distances, requires a large number of elements or a very large RIS. This is primarily because the RIS-aided link suffers from double path loss, which is accordingly referred to as product-distance path loss~\cite{wu2021}.

\begin{figure}
    \centering
    \subfloat[Number of elements for $\rho_t=20\,\text{m}$]{\includegraphics[width=\columnwidth]{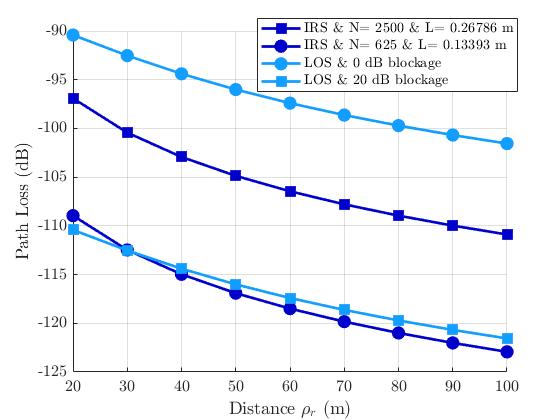}}
    \hfill
    \subfloat[RIS length for $\rho_t=\rho_r $]{\includegraphics[width=\columnwidth]{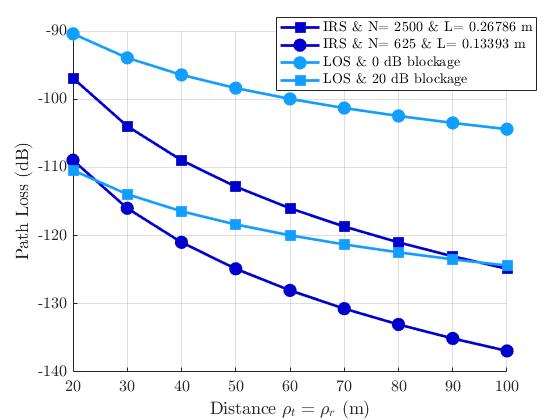}}
    \caption{Path loss against distance at 28 GHz.}
    \label{fig:hhi:f10}
\end{figure}

\begin{figure}
    \centering
    \subfloat[Number of elements]{\includegraphics[width=\columnwidth]{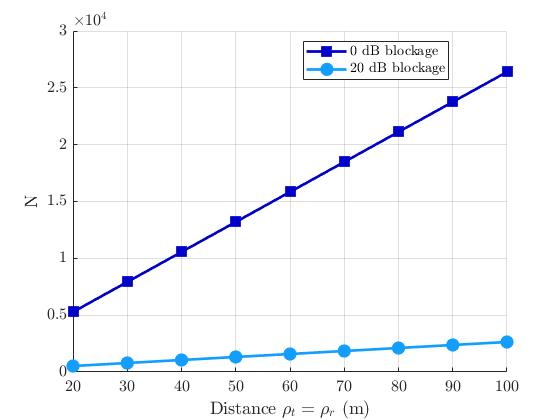}}
    \hfill
    \subfloat[RIS length]{\includegraphics[width=\columnwidth]{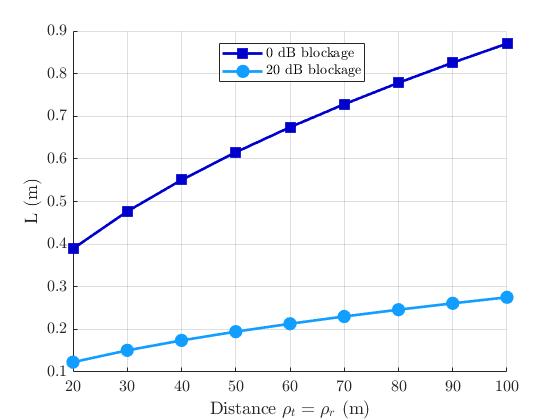}}
    \caption{Required number of RIS elements and required RIS size against distance at 28 GHz.}
    \label{fig:hhi:f11}
\end{figure}

Figures~\ref{fig:hhi:f10} and~\ref{fig:hhi:f11} correspond to Figures~\ref{fig:hhi:f8} and~\ref{fig:hhi:f9}, respectively, but for a frequency of 28 GHz. The trends are generally consistent with the 6 GHz calculations. In addition, by comparison, it can be seen that the signals generally experience greater attenuation at higher frequencies. Due to the product-distance path loss, far more elements are then required ($625$ elements are not sufficient here to compensate for the path loss of the LOS link with 20 dB blockage, but $2500$ elements are) than at a frequency of 6 GHz. At the same time, this can be achieved with smaller RISs (approximately $L=0.26786$ m), since the size of the elements and thus the overall size of the RIS is frequency-dependent, e.g., assuming an element size of $\lambda/2$, where $\lambda$ corresponds to the RIS wavelength of operation.


\section{Discussion}\label{sec:disc}

This section discusses the results presented in Section~\ref{sec:hw} about hardware and in Section~\ref{sec:depl} about deployment. It is intentional that only the first steps and ideas of the 6G-LICRIS project are presented. The results will be presented in dedicated future publications regarding the hardware realization with LC, the propagation modeling, and system architecture aspects, as well as the deployment optimization, especially with the use and description of the channel model QuaDRiGa including RIS.

In Section~\ref{sec:hw} about the hardware, the RIS topology was presented where the realization was with liquid crystal. In most cases, RISs are realized with semiconductor-based approaches, commonly with PIN diodes or varactor diodes, and only a relatively few works exist with the realization of RIS with liquid crystal~\cite{rana23,neuder24,matsuno22kddi}. Compared to semiconductor-based RIS, LC is more energy-efficient in higher frequencies, it is scalable, cost-effective, and a promising technology candidate for RIS hardware implementation.

Two different types of RIS were presented: The first one is the multi-resonance RIS. The novelty of it is the large wideband, since most designs have a relative bandwidth of around 10\%, the presented design boasts more than 20\% bandwidth. It is also compatible with very thin LC-layers, which makes the tuning time of the LC much faster and reduces the amount of LC needed, and therefore, is capable of reducing costs.

Regarding the delay-line based RIS, it is comparable, for example, with Reference~\cite{neuder24}, but these approaches are based on one single polarization only. Theoretically, this approach should be compatible with dual polarization by placing two perpendicular lines. However, in order to obtain the required phase shift, the delay-lines need to be quite long, and it is not possible to place two lines in the same unit-cell. What is proposed is to significantly reduce the dimension that the delay-lines need to obtain the phase difference, and therefore, the required space to place two lines is achieved. Unfortunately, this also implies an increase in losses.

In Section~\ref{sec:propmodel}, the propagation modeling was presented. This suits the integration of a RIS into a mobile network. A RIS integration mechanism with the existing 3GPP network architecture is proposed in the following. The approach for this is to use, for example, an open radio access network (O-RAN) architecture~\cite{polese23} to include the RIS orchestrator and controller. The key components within the O-RAN architecture are the Radio Unit (RU) domain and the Central Unit (CU) domain, with an additional intelligent controller (IC) function. In addition, the non-real-time radio access network intelligent controller (Non-RT RIC) logical function resides within the Service Management and Orchestration (SMO) framework and provides non-real-time optimization and control of RAN resources. It facilitates functionalities like artificial intelligence/machine learning (AI/ML) model training and policy-based management for near-real-time RIC applications. The RIS is now integrated regarding RIS orchestration in the Non-RT RIC and regarding RIS control in the near-real-time RIC. The advantage of this approach is that no new interfaces are needed. For the AI/ML model, approaches as are used. A detailed description of such models, as well as the architecture including RIS, is planned in future publications.

A deployment optimization was presented in Section~\ref{sec:deploptim} as well. A channel model is used to calculate the coverage in an urban scenario for the system consisting of BS-UE and BS-RIS-UE compared to BS-NCR-UE~\cite{aghazadeh23hua}. The coverage is shown as a heatmap of the SNR. The findings suggest that a RIS is suitable for open spaces such as squares, while the NCR shows better improvements for narrow scenarios such as streets. In the present publication, only the RIS without NCR consideration for 2-dimensional scenarios is presented. Also, only the improved area is shown without a detailed signal strength or heatmap. However, the present publication demonstrated different RIS locations to optimize the deployment, which is missing in~\cite{aghazadeh23hua}. There are even deep reinforcement learning/artificial intelligence approaches for the deployment optimization, however, only for indoor applications~\cite{encinaslago23nec}. Outdoor measurements of a RIS at $26$ GHz, constructed with PIN diodes and equipped with $1024$ elements and a predefined codebook, are presented in reference~\cite{liu22zte}. They provide a coverage analysis both with and without the RIS. The outdoor measurement only should be compared with the planned future publications. 

The RIS dimensioning was presented in Section~\ref{sec:dimensioning}. This was an important consideration for the specification and the RIS size, which was also considered in Section~\ref{sec:hw}. One of the main findings is that a high number of RIS elements, $N$, is helpful to compensate for the low channel gain. This is in accordance with, for example, Reference~\cite{bjornson2019} (see also Section~\ref{sec:ncr}).

Recently, new technologies that extend the RIS concept have emerged. For example, stacked intelligent metasurfaces adopt a different topology to perform computations in the wave domain, allowing for more degrees of freedom in signal processing~\cite{an23,an24no3}. While the concept is conceptually promising, it might bring several engineering challenges that hinder its adoption in commercial products. For example, the stacked design incurs additional transmission losses that are not acceptable in real products. Furthermore, the meta-atoms might introduce non-linearity distortion in the system that complicates conformance to regulation and out-of-band emissions. In any case, these issues are open problems that remain to be solved in future work.

\section{Conclusions and Outlook}
This paper presents an overview of RIS technology from an industry R\&D perspective under the scope of the 6G-LICRIS project. Potential use cases, technology requirements, and existing challenges are outlined. RIS hardware concepts are presented, and practical problems such as RIS placement and dimensioning, as well as radio propagation modeling, are discussed. A hardware design of a RIS with liquid crystal at 28 GHz is presented. A propagation model for RIS as a new part of the system architecture is discussed with the approaches of semi-empirical models, geometric models, and their combination through the application of artificial intelligence/machine learning. Finally, a channel model for deployment optimization and dimensioning is presented, with the findings that a rather large RIS is favored for coverage improvement as well as greater attenuation at higher frequencies combined with a smaller RIS size. The 6G-LICRIS project results will be presented in dedicated future publications regarding the hardware realization with LC, the propagation modeling, and system architecture aspects, as well as the deployment optimization, especially with the use and description of the channel model QuaDRiGa including RIS.

Overall, while RIS is a promising technology for 6G, various practical problems need to be addressed before incorporating such a solution into mobile cellular systems. In particular, RIS channel estimation, control signaling overhead, and coexistence concerns are some of the serious challenges that remain largely unsolved and shall be addressed in future work.

\bibliographystyle{IEEEtran}
\bibliography{references}


\end{document}